\documentclass[12pt]{article}
\usepackage{color}
\usepackage{epsfig}
\usepackage{amsfonts}
\usepackage{amsmath}
\usepackage{amssymb}
\usepackage{bm}
\usepackage{cite}
\usepackage{rotating}
\usepackage{array}
\usepackage{multirow}
\usepackage[title]{appendix}
\begin{document}
\sffamily
\title{
Jost-matrix analysis of the resonance
${}^5\mathrm{He}^*(\frac32^+)$ near the $dt$-threshold
}
\author{
S.A. Rakityansky$^{\,a}$\footnote{e-mail: rakitsa@up.ac.za}\quad
and\quad
S.N. Ershov$^{\,b}$\footnote{e-mail: ershov@theor.jinr.ru}\\[3mm]
\parbox{11cm}{%
${}^a${\small Department of Physics, University of Pretoria, Pretoria,
South Africa}\\
${}^b${\small Joint Institute for Nuclear Research, Dubna, Russia}
}
}
\maketitle
\begin{abstract}
\noindent
Experimental data on the $n\alpha$ and $dt$ collisions in the quantum state 
$J^\pi=\frac32^+$ near the $dt$-threshold are fitted using the semi-analytic 
multi-channel Jost matrix with proper analytic structure and some adjustable 
parameters. Then the spectral points are sought as zeros of the Jost matrix 
determinant (which correspond to the $S$-matrix poles) at complex energies. The 
correct analytic structure makes it possible to calculate the fitted Jost matrix 
on any sheet of the Riemann surface whose topology involves not only the 
square-root but also the logarithmic branching caused by the Coulomb 
interaction. Within a distance of 100\,keV above the $dt$-threshold, three 
$\frac32^+$-resonances are found on the non-physical sheet of the Riemann 
surface. Several $S$-matrix (shadow) poles on the other sheets of this surface 
are located as well.
\end{abstract}

\section{Introduction}
The five-nucleon system ${}^5\mathrm{He}$ is unstable even in its ``ground
state'', which actually is a resonance lying 798\,keV above the
$n\alpha$-threshold\cite{Tilley}. The other decay channel, $dt$, opens up
much higher, at 16.792\,MeV above the ground state\cite{Tilley}.

This system plays an important role in nuclear astrophysics as well as in the
man-made thermonuclear fusion (for both the peaceful and military purposes).
This is why so many publications have been devoted to experimental and
theoretical studies of its properties
(see, for example, few selected papers
\cite{
Haesner,
Jarmie,
Brown,
PRL59,
Karnakov,
Bogdanova,
CsotoLovas,
Efros,
CsotoHale,
Barker,
Drosg,
Hoop,
Navratil,
BrownHale,
HaleBrown,
Hoop15,
betan}).
One of the controversial issues in
these studies, that still remains vague, is the nature of the resonant state
with $J^\pi=\frac32^+$ lying at around $\sim50$\,keV above the $dt$-threshold.
The majority of theoretical analyses claim that this state is a superposition
of a conventional resonance and a so-called ``shadow''
$S$-matrix pole that supposedly is located somewhere on one of the
non-physical sheets of the energy Riemann-surface and somehow affects from there
the physical collisions.

The main task of our present paper is to clarify the nature of that
${}^5\mathrm{He}^*(\frac32^+)$-resonance.
As the experimental data, we use the partial cross sections for the four
coupled channels, $n\alpha(\ell=2,s=1/2)$, $dt(\ell=0,s=3/2)$,
$dt(\ell=2,s=1/2)$, and $dt(\ell=2,s=3/2)$, that we obtain using the $R$-matrix
parametrization given in Ref.~\cite{PRL59}. A closer look at the
inter-channel cross sections leads us to
the conclusion that only the first two of them, namely, $n\alpha(\ell=2,s=1/2)$
and $dt(\ell=0,s=3/2)$, are strongly coupled to each other. It is therefore
reasonable to describe the $\frac32^+$-resonances near the $dt$-threshold within
the corresponding two-channel model. This is what we do in the present paper.

The novelty of our analysis consists in using the multi-channel Jost matrix for 
fitting the partial cross sections. These cross sections are extracted from 
the available $R$-matrix parametrization of a large collection of experimental 
data. Therefore, in this way, we indirectly fit the data.
After the fitting at real energies, the Jost matrix is considered at complex 
$E$, where the zeros of its determinant correspond to the poles of the 
$S$-matrix.

We use a special representation of the milti-channel Jost matrix suggested in
Refs.~\cite{our.MultiCh, our_Coulomb}, where the Jost matrix is given as a sum
of two terms and each term is factorized in a product of two matrices. One
of these matrices is an unknown analytic single-valued function of the energy
and the other is given explicitly as a function of the channel momenta and is
responsible for the branching of the Riemann surface. The unknown single-valued
matrices are parametrized and the parameters are found via fitting the
available experimental data.

When using the semi-analytic representation of the Jost matrix, where the
factors responsible for the topology of the Riemann surface are given
explicitly, it is easy to explore the behaviour of the Jost matrix on all the
sheets of the Riemann surface. In this way we are able to accurately locate
the resonance poles and to examine the existence of the ``shadow'' poles.

\section{Jost matrices}
In this Section, we give a definition of the Jost matrices, show how they are
related to the observable quantities, and discuss their analytic properties.
All the details omited here, can be found in
Refs.~\cite{our.MultiCh,our_Coulomb,two_channel}.

\subsection{Definition}
We restrict our consideration to the binary multi-channel
reactions. In such a reaction the colliding particles may either change their
internal states ($a+b\to a^*+b^*$) or transform themselves into another pair of
particles ($a+b\to c+d$). Generally speaking, the masses of the particles (and
therefore the total energy) in the initial and final state are different and
thus the channels $ab$, $a^*b^*$, $cd$, etc. have different thresholds. Each of
these different two-particle states is labeled by the symbol $\gamma$. For a
given $\gamma$, the particles may have several possible combinations of the
orbital angular momentum, $\ell$ and
the total spin $s$. Each combination of them (despite the fact that the
threshold energy is the same) is treated as a separate channel. The total
angular momentum $J$ and the parity $\pi$ are the same for all the channels and,
in order to simplify the notation, we omit them. Therefore a channel is
specified by the three quantum
numbers, $(\gamma, \ell, s)$. Each possible combination of the numbers $(\gamma,
\ell, s)$ can be assigned a single sequential number $n=1,2,\dots,N$, where $N$
is the total number of the channels taken into account in a particular model.
For example, in the problem we consider in the present paper, namely, the
quasi-bound state of ${}^5\mathrm{He}$ with $J^\pi=\frac32^+$ near the
$dt$-threshold, there are
four channels listed in Table~\ref{table.channels}.

\begin{table}
\begin{center}
\sffamily
\begin{tabular}{|c|c|c|c|c|c|}
\hline
   & $\gamma$ & $n\alpha$ & \multicolumn{3}{c|}{$dt$}\\
\cline{2-6}
 ${}^5\mathrm{He}^*(\frac32^+)$ & $(\ell, s)$ & $(2,\frac12)$ & $(0,\frac32)$
& $(2,\frac12)$ & $(2,\frac32)$\\
\cline{2-6}
 & $n$ & 1 & 2 & 3 & 4\\
\hline
\multicolumn{2}{|c|}{barriers}&
centrifugal & Coulomb &
\parbox{2.1cm}{centrifugal${\phantom{\frac12}}^{\mathstrut}$\\+ Coulomb}&
\parbox{2.1cm}{centrifugal${\phantom{\frac12}}^{\mathstrut}$\\+ Coulomb}\\[3mm]
\hline
\multicolumn{2}{|c|}{\parbox{2.8cm}{relative\\kinetic energies}}&
$\sim18$\,MeV & $\sim50$\,keV  & $\sim50$\,keV & $\sim50$\,keV \\
\hline
\end{tabular}
\end{center}
\caption{\sf
Four possible combinations of the set $(\gamma, \ell, s)$ for the $\frac32^+$
state of ${}^5\mathrm{He}$ just above the $dt$-threshold; types of the
potential barriers in each of the channels; and the relative kinetic
energies of the fragments in the resonance state.
}
\label{table.channels}
\end{table}

The system of the $N$-channel radial Schr\"odinger equations,
\begin{eqnarray}
\label{coupled_radial}
    \left[\partial^2_r+k_n^2-\frac{\ell_n(\ell_n+1)}{r^2}
    -\frac{2k_n\eta_n}{r}\right]
    u_{n}(E,r)
    =
    \sum_{n'=1}^NV_{nn'}(r)u_{n'}(E,r)&,&\\[3mm]
\nonumber
    n = 1,2,\dots,N &,&
\end{eqnarray}
where $k_n$, $\ell_n$, $\eta_n$, and $V$ are the channel momentum,
channel angular momentum, channel Sommerfeld parameter, and the
short-range part of the potential, has $2N$ linearly independent column
solutions, $(u_1,u_2,\dots,u_N)^T$. Since this differential equation has a
regular singular point at $r=0$, only half of these solutions are regular at
$r=0$ (see, for example, Ref.~\cite{Riley}). These $N$ regular columns form a
regular basis, i.e. any other
regular solution (the physical solution in particular) is their linear
combination. Putting  these regular columns together, we obtain a square
($N\times N$)-matrix $\boldsymbol{\phi}(E,r)$. At large distances the right-hand side
of Eq.~(\ref{coupled_radial}) vanishes and it becomes pure Coulomb equation.
This means that, when $r\to\infty$, each column of matrix $\bm{\phi}(E,r)$
becomes a linear combination of the pure Coulomb functions. For the case of
$N=2$ this can be written as
\begin{eqnarray}
\nonumber
  \bm{\phi}(E,r)
  &\mathop{\longrightarrow}\limits_{r\to\infty}&
  \begin{bmatrix}
  H_{\ell_1}^{(-)}(\eta_1,k_1r)e^{i\sigma_{\ell_1}} & 0\\[3mm]
  0 & H_{\ell_2}^{(-)}(\eta_2,k_2r)e^{i\sigma_{\ell_2}}
  \end{bmatrix}
  \begin{bmatrix}
  f_{11}^{\mathrm{(in)}}(E) & f_{12}^{\mathrm{(in)}}(E)\\[3mm]
  f_{21}^{\mathrm{(in)}}(E) & f_{22}^{\mathrm{(in)}}(E)
  \end{bmatrix}
  +\\[3mm]
\label{regular_ass}
  &+&
  \begin{bmatrix}
  H_{\ell_1}^{(+)}(\eta_1,k_1r)e^{-i\sigma_{\ell_1}} & 0\\[3mm]
  0 & H_{\ell_2}^{(+)}(\eta_2,k_2r)e^{-i\sigma_{\ell_2}}
  \end{bmatrix}
  \begin{bmatrix}
  f_{11}^{\mathrm{(out)}}(E) & f_{12}^{\mathrm{(in)}}(E)\\[3mm]
  f_{21}^{\mathrm{(out)}}(E) & f_{22}^{\mathrm{(in)}}(E)
  \end{bmatrix}\ ,
\end{eqnarray}
where
\begin{equation}
\label{Riccati_Coulomb}
   H_\ell^{(\pm)}(\eta,kr)=F_\ell(\eta,kr)\mp iG_\ell(\eta,kr)
   \ \mathop{\longrightarrow}\limits_{r\to\infty}
   \ \mp i\exp\left\{\pm i\left[kr-\eta\ln (2kr)
   -\frac{\ell\pi}{2}+\sigma_\ell\right]\right\}
\end{equation}
are the in-coming and out-going Coulomb spherical waves, $\sigma_\ell(E)$ is the
pure Coulomb phase shift, and the combination coefficients are organized in the
($N\times N$)-matrices $\bm{f}^{\mathrm{(in/out)}}(E)$ that are called the Jost
matrices. In a sense, they are the amplitudes of the incoming and outgoing waves
in the asymptotic behaviour of the regular solution. Of course the matrices
$\bm{f}^{\mathrm{(in)}}$ and $\bm{f}^{\mathrm{(out)}}$ are not completely independent. At
real collision energies they are complex conjugate of each other. When $E$ is
complex, the relation between them becomes more complicated. Their values on
different sheets of the Riemann surface are related as is shown in the
Appendix~\ref{Appendix_symmetry}.

\subsection{Observables}
Since the columns of matrix $\bm{\phi}(E,r)$ constitute a regular basis, a physical
wave function, i.e. a column $\bm{u}(E,r)$, is their linear combination (for the
sake of clarity, we write the formulae for $N=2$):
\begin{eqnarray}
\nonumber
   \bm{u}(E,r) &=& \begin{bmatrix} u_1(E,r)\\ u_2(E,r)\end{bmatrix}=
   \begin{bmatrix} \phi_{11}(E,r)\\ \phi_{21}(E,r)\end{bmatrix}c_1+
   \begin{bmatrix} \phi_{12}(E,r)\\ \phi_{22}(E,r)\end{bmatrix}c_2\\[3mm]
\label{phys_comb}
   &=&
   \begin{bmatrix} \phi_{11}(E,r) & \phi_{12}(E,r)\\
			 \phi_{21}(E,r) & \phi_{22}(E,r)\end{bmatrix}
   \begin{pmatrix} c_1 \\ c_2 \end{pmatrix}
   =\bm{\phi}(E,r)\bm{c}\ .
\end{eqnarray}
The combination coefficients $c_n$ are to be chosen to satisfy certain physical
boundary conditions at infinity. For a spectral point (either bound or a
resonant state) the physical wave function should only have the outgoing waves
in its asymptotic behaviour,
\begin{eqnarray}
\nonumber
  \bm{u}(E,r)
  &\mathop{\longrightarrow}\limits_{r\to\infty}&
  \begin{bmatrix}
  H_{\ell_1}^{(-)}(\eta_1,k_1r)e^{i\sigma_{\ell_1}} & 0\\[3mm]
  0 & H_{\ell_2}^{(-)}(\eta_2,k_2r)e^{i\sigma_{\ell_2}}
  \end{bmatrix}
  \bm{f}^{\mathrm{(in)}}(E)\bm{c}
  +\\[3mm]
\label{spectral_bc}
  &+&
  \begin{bmatrix}
  H_{\ell_1}^{(+)}(\eta_1,k_1r)e^{-i\sigma_{\ell_1}} & 0\\[3mm]
  0 & H_{\ell_2}^{(+)}(\eta_2,k_2r)e^{-i\sigma_{\ell_2}}
  \end{bmatrix}
  \bm{f}^{\mathrm{(out)}}(E)\bm{c}\ .
\end{eqnarray}
This can only be achieved if
\begin{equation}
\label{fczero}
  \bm{f}^{\mathrm{(in)}}(E)\bm{c}=
  \begin{bmatrix}
  f_{11}^{\mathrm{(in)}}(E) & f_{12}^{\mathrm{(in)}}(E)\\[3mm]
  f_{21}^{\mathrm{(in)}}(E) & f_{22}^{\mathrm{(in)}}(E)
  \end{bmatrix}
  \begin{pmatrix} c_1 \\ c_2 \end{pmatrix}=0\ ,
\end{equation}
which is a homogeneous system of linear equations for the unknown combination
coefficients $c_n$. It has a non-zero solution if and only if
\begin{equation}
\label{detfzero}
  \det
  \begin{bmatrix}
  f_{11}^{\mathrm{(in)}}(E) & f_{12}^{\mathrm{(in)}}(E)\\[3mm]
  f_{21}^{\mathrm{(in)}}(E) & f_{22}^{\mathrm{(in)}}(E)
  \end{bmatrix}
  =0\ .
\end{equation}
The roots $E=\mathcal{E}_n$ of this equation at real negative energies
($\mathcal{E}_n<0$)
correspond to the bound states, and the roots at complex energies
($\mathcal{E}_n=E_r-i\Gamma/2$) give us the resonances.

In a similar way it can be easily shown (see, for example,
Refs.~\cite{our.MultiCh,two_channel}) that the scattering is determined by the
``ratio'' of the amplitudes of the out-going and in-coming waves, i.e. by the
$S$-matrix,
\begin{equation}
\label{Smatrix}
   \bm{S}(E)=\bm{f}^{\mathrm{(out)}}(E)
   \left[\bm{f}^{\mathrm{(in)}}(E)\right]^{-1}\ .
\end{equation}
It is obvious that the roots of eq.~(\ref{detfzero}) correspond to the poles of
the $S$-matrix.
It can be shown (see, for example, Ref.~\cite{Frobrich}) that the
reaction cross section is determined by the reaction amplitude,
\begin{equation}
\label{multichannel.partial.Amplitude}
   \mbox{\sl f}^{\,J}_{n'\gets n}=
   \frac{S^J_{n'n}-\delta_{n'n}}{2ik_n}i^{\ell_n-\ell'_{n'}}\ ,
\end{equation}
that is expressed via the corresponding $S$-matrix.
The partial cross section (describing the transition between any two particular
channels) can be obtained as
\begin{equation}
\label{particular_sigma}
   \sigma^J(\gamma'\ell's'\gets\gamma\ell s)=4\pi
   \frac{\mu_\gamma k_{\gamma'}}{\mu_{\gamma'}k_\gamma}\cdot
   \frac{2J+1}{2s+1}\left|\mbox{\sl f}^{\,\,J}_{\,n'n}\right|^2\ ,
\end{equation}
where $\mu_\gamma$ is the reduced mass corresponding to the state $\gamma$
and the subscript $n$ is a sequential number that labels all possible
combinations of the quantum numbers $\{\gamma,\ell,s\}$ (see
Table~\ref{table.channels}).

If we do not monitor the spin states, then the total cross section for the
transition from the initial internal state $\gamma$ of the colliding particles
to the final state $\gamma'$ (the final particles can be different from the
initial ones) is given by
\begin{equation}
\label{multichannel.sigma.etaprimeeta}
   \sigma(\gamma'\gets\gamma)=
   \sum_{J\ell's'\ell s}\sigma^J(\gamma'\ell's'\gets\gamma\ell s)\ .
\end{equation}

\subsection{Analytic properties}
The Jost matrices (and thus the $S$-matrix) are not single-valued functions of
the energy. There are two reasons for this:
\begin{itemize}
\item
The in-coming and out-going spherical waves (\ref{Riccati_Coulomb}), and thus
their amplitudes, $\bm{f}^{\mathrm{(in/out)}}(E)$, depend on $E$ via all the channel
momenta;
\item
For charged particles, there is an additional complication, namely, the
in-coming and out-going spherical waves (and thus their amplitudes) depend on
$\ln(k_n)$.
\end{itemize}
For the channel momenta,
\begin{equation}
\label{ch_momenta}
   k_n=\pm\sqrt{\frac{2\mu_n}{\hbar^2}(E-E_n)}\ ,
   \qquad  n=1,2,\dots,N\ ,
\end{equation}
where $E_n$ are the threshold energies, there are $2^N$ possible
combinations of the signs in front of the $N$ square roots (\ref{ch_momenta}).
Therefore for each value of the energy the Jost matrix has $2^N$ different
values. 

The complex function $\ln(k_n)$ has infinitely many different values,
\begin{equation}
\label{Logarithm_ch_momenta}
   \ln(k_n)=\ln\left\{|k_n|e^{i[\arg(k_n)+2\pi m]}\right\}=
   \ln|k_n|+i\arg(k_n)+i2\pi m\ ,
   \qquad  m=0,\pm1,\pm2,\dots\ ,
\end{equation}
corresponding to different choices of $m$. It is defined on the ``spiral'' 
Riemann surface with the branch point at $k_n=0$. It is customary to define the 
so called principal branch of the logarithm as the part of this surface 
corresponding to $m=0$ and $-\pi<\arg(k_n)\leqslant\pi$. The value of the 
function $\ln(k_n)$ in Eq.~(\ref{Logarithm_ch_momenta}) corresponds to the 
energy
\begin{equation}
\label{E4pi}
      E-E_n=|E-E_n|\exp(i\varphi_n+i4\pi m)\ ,
\end{equation}
where $-2\pi<\varphi_n\leqslant 2\pi$ and $\arg(k_n)=\varphi_n/2$. Since the 
momentum is the square root of the energy, the interval 
$-2\pi<\varphi_n\leqslant 0$ covers the unphysical (with respect to channel $n$) 
Riemann sheet where $\mathrm{Im}\,k_n<0$, and the interval $0<\varphi_n\leqslant 
2\pi$ corresponds to the physical sheet with $\mathrm{Im}\,k_n<0$. These two 
intervals are represented by the two sheets of the Riemann surface stemming from 
the square-root branch point. Generally speaking, each thereshold energy, $E_n$, 
is a square-root and a logarithmic branch point at the same time.

In order to have a one-to-one correspondence between the energy and the set of 
the channel momenta and their logarithms, the multiple copies (sheets) of the 
energy plane are 
used. For each sheet the imaginary parts of all the channel momenta as well as 
the corresponding logarithmic indices $m$ have definite values. Each of these 
sheets is cut along its own real axis from the lowest threshold to infinity. The 
edges of the cuts of different sheets are interconnected in such a way that the 
path around one or more thresholds leads to an appropriate changes of 
$\mathrm{Im}\,k_n$ and $m$. On the multi-layered Riemann surface obtained in 
this way, the Jost matrices (and therefore the $S$-matrix) are single-valued 
functions of the energy.

In the ``simple'' case, when the Coulomb potential is absent, the Jost matrices
are defined on a Riemann surface consisting of $2^N$ sheets.
 Apparently, the sign of $\mathrm{Im}\,k_n$ is determined by the choice of the 
sign in Eq.~(\ref{ch_momenta}). The points sitting on a vertical line passing 
through all the sheets, correspond to the same energy but to different choices 
of the signs in Eqs.~(\ref{ch_momenta}).

\begin{figure}
\centerline{\epsfig{file=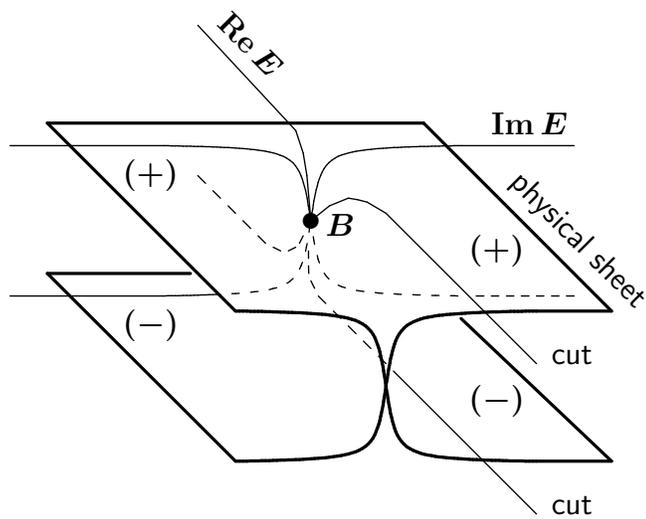}}
\caption{\sf
Fragments of the physical $(+)$ and unphysical $(-)$ sheets of a
single-channel Riemann surface around the branch point ($B$) that
corresponds to the threshold energy $E_1=0$. Each sheet has its own real and
imaginary axes, and the point $B$ is their common zero-point.
Transition from $(+)$ to $(-)$ and back is possible through the
cuts of both sheets, running from $B$ to infinity along positive half of the
corresponding real axes.
}
\label{fig.1ch_Riemann}
\end{figure}

At each threshold (where $k_n=0$ and therefore the
choice of the sign for $k_n$ is immaterial) there are $2^{N-1}$ pairs of
Riemann sheets that touch each other. Moving around such a branch point, we
pass from one sheet to the other and back, because the sign of the
corresponding channel momentum changes to the opposite after one full circle.
For a single-channel case ($N=1$), such a Riemann surface is schematically
shown in Fig.~\ref{fig.1ch_Riemann}.

\begin{figure}
\centerline{\epsfig{file=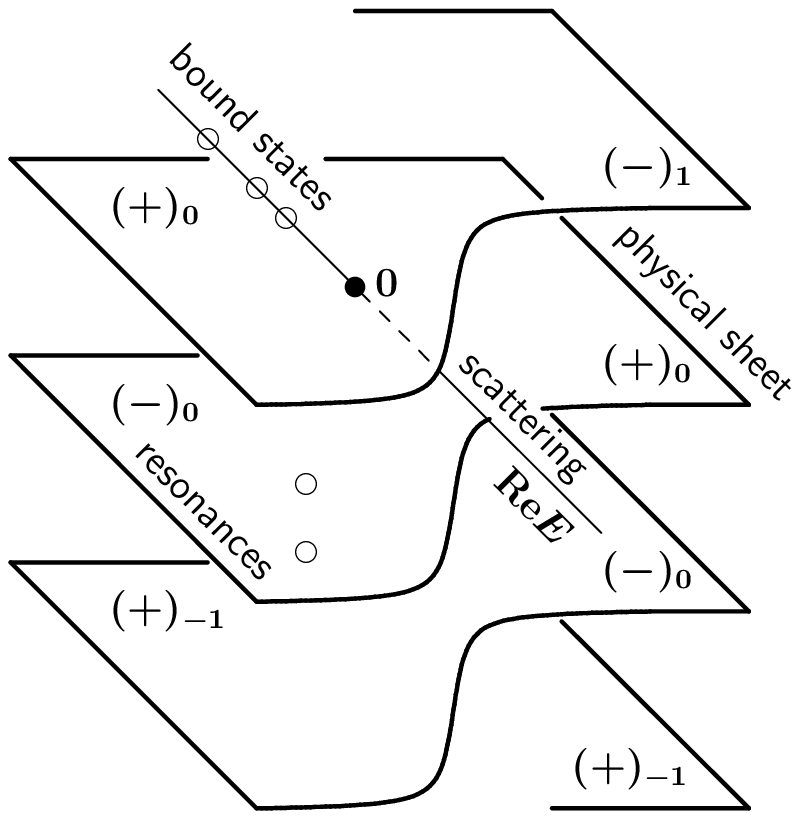}}
\caption{\sf
Spiral Riemann surface for a single-channel potential with a Coulomb tail. The
symbols $(\pm)_m$ label the sheets where $\mathrm{Im}(k)$ is either positive or
negative, and the subscript $m$ is the number of $i2\pi$ in
Eq.~(\protect{\ref{Logarithm_ch_momenta}}) stemming from the logarithmic
branching.
}
\label{fig.single.Coulomb.Riemann}
\end{figure}

When a Coulomb potential is present, the first full circle around the branch 
point changes the sign of $\mathrm{Im}\,k_n$, while the second full circle 
brings back the original sign but also increases (anti-clockwise direction) or 
decreases (clockwise direction) $m$ in Eq.~(\ref{Logarithm_ch_momenta}) by one 
unit. This means that moving in the same direction (clockwise or 
anti-clockwise), we can never come back to the same sheet. For the 
single-channel case, this is illustrated in 
Fig.~\ref{fig.single.Coulomb.Riemann}.

The sheet of the Riemann surface where $\mathrm{Im}\,k_n>0$ is called physical 
sheet with respect to the channel $n$. Otherwise it is called non-physical with 
respect to that channel. The scattering energies are on the upper rim of the cut 
of the sheet that is physical with respect to all the channels. The resonances 
are the discrete points (where the $S$-matrix is singular) on the sheet that is 
non-physical with respect to all open channels. This is because the open-channel 
components of the resonance wave functions asymptotically behave as pure 
outgoing divergent spherical waves, and this is only possible if 
$\mathrm{Im}\,k_n<0$ for all $n$ corresponding to the open channels.

In addition to the bound-state and resonance spectral points, the multi-channel 
$S$-matrix may have poles at ``wrong'' locations on the Riemann surface, i.e. 
where the corresponding wave function asymptotically does not behave 
``correctly''. Such solutions of the Schr\"odinger equation have no physical 
meaning. However, mathematically these ``wrong'' poles may influence the 
behaviour of the $S$-matrix at the physical scattering energies, if they are 
close enough to the real axis passing through the physical sheet. Such singular 
points are called the ``shadow'' poles of the $S$-matrix. Apparently, if such a 
shadow pole has an influence on the observable quantities, it cannot be ignored, 
but (in contrast to the resonances) the imaginary part of the corresponding 
energy cannot be interpreted as a width of a decay process (because the 
corresponding wave function has wrong asymptotic behaviour).

As is seen, the Coulomb forces drastically change the Riemann surface. However,
we can safely ignore all the sheets with $m\neq0$, because any possible
$S$-matrix poles on them are far away from the upper rim of the real axis on
the physical sheet $(+)_0$, where the scattering takes place, and thus such
poles would have no influence. The sheets with $m=0$ correspond to the
so-called principal branch of the logarithm function and therefore can be called
principal sheets of the Riemann surface. The only thing that should not be
forgotten is that, in contrast to Fig.~\ref{fig.1ch_Riemann}, it is not possible
to pass from $(+)_0$ to $(-)_0$ when moving in the anti-clockwise direction. In
other words, with the presence of a Coulomb potential the principal sheets
of the Riemann surface are not all interconnected among themselves.

\begin{figure}
\centerline{\epsfig{file=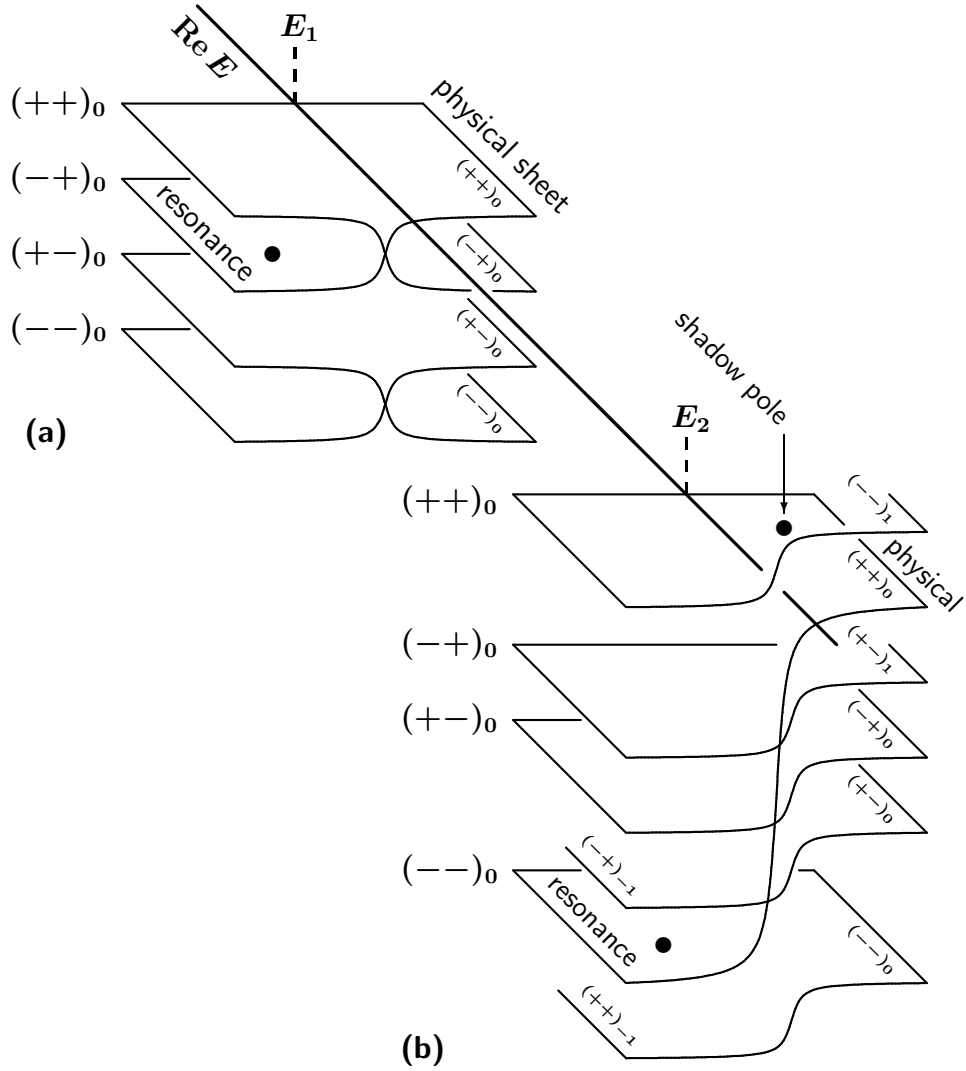}}
\caption{\sf
Interconnections of the Riemann sheets for the two-channel problem, where the
Coulomb potential is only present in the second channel:
(a) the interconnections between the thresholds $E_1$ and $E_2$; (b) the
interconnections above the highest threshold. The symbols
$(\pm\pm)_m$ label the sheets where $\mathrm{Im}(k_1)$ and $\mathrm{Im}(k_2)$
are either positive or
negative, and the subscript $m$ is the number of $i2\pi$ in
Eq.~(\protect{\ref{Logarithm_ch_momenta}}) for the second channel.
}
\label{fig.sheets_2ch_Coulomb}
\end{figure}

In the present paper we deal with a two-channel problem, where there is no
Coulomb potential in the first channel ($n\alpha$), but it is present in the
second one ($dt$). The corresponding Riemann surface is shown in
Fig.~\ref{fig.sheets_2ch_Coulomb}.
There are four ($2^N$) principal sheets, namely, $(++)_{0}$, $(+-)_{0}$,
$(-+)_{0}$, and $(--)_{0}$, where the subscript indicates the value of $m$ in
Eq.~(\ref{Logarithm_ch_momenta}) for the second channel. The cuts between $E_1$
and $E_2$ are interconnected as for the neutral particles. The logarithmic
branch point is at
the second threshold, beyond which the sheets have the spiral interconnections.
Of course there are infinite number of the sheets at any energy. For the low
energies, they are the direct extensions of all the spiral sheets shown for
$E>E_2$. However, in order to reach them from the low energies, we have to go
around the branch point at $E_2$. As was already mentioned, we can ignore them
because the behaviour of the Jost matrices on them cannot affect any observable
quantities.

\subsection{Analytic structure}
It is seen that even in the simple two-channel case the Riemann surface has a
very complicated topology. Therefore when doing an analytic continuation of the
Jost matrix (or the $S$-matrix) from the real axis to complex energies, one
should be careful. This is especially true when such a continuation is done
near a branch point (i.e. near a threshold energy).

Very often the $S$-matrix is defined with additional square
roots of the channel momenta (see, for example, Ref.~\cite{Taylor}),
\begin{equation}
\label{SmatrixTilde}
   \tilde{S}_{n'n}=\sqrt{k_{n'}}\sum_{n''}f^{\mathrm{(out)}}_{n'n''}
   \left[f^{\mathrm{(in)}}\right]^{-1}_{n''n}\frac{1}{\sqrt{k_n}}\ ,
\end{equation}
and with the corresponding re-definition of the reaction amplitude
(\ref{multichannel.partial.Amplitude}),
\begin{equation}
\label{Amplitude_tilde}
   {\mbox{\!{\sl f}}}^{\,\,J}_{n'\gets n}=
   \frac{\tilde{S}^J_{n'n}-\delta_{n'n}}
   {2i\sqrt{k_{n'}k_n}}i^{\ell_n-\ell'_{n'}}\ .
\end{equation}
Such a definition makes the $S$-matrix symmetric and unitary on the real axis.
This is convenient, but the price one has to pay for such a convenience is the
uncertainty when matrix $\tilde{S}(E)$ is continued to complex energies in
search for resonances. Indeed, the root of the fourth order,
$\sqrt{k_n}\sim\sqrt[4]{E-E_n}$, of a complex number has four different values.
For example $\sqrt[4]{1}$ is equal either to $\pm1$ or to $\pm i$.
In other words, each threshold $E_n$ becomes a fourth-order
branch point ($\sqrt[4]{E-E_n}$) instead of the second order
($\sqrt[2]{E-E_n}$). As a result the number of the Riemann sheets is
artificially increased. The
same difficulty shows up when the so called $K$-matrix is used for the analytic
continuation. We do not claim that using the matrix (\ref{SmatrixTilde}) always
leads to wrong analytic continuation, but one should be extremely careful,
especially when doing numerical calculations because different compilers work
with the complex numbers and with multi-valued functions differently.

In our approach, neither Eq.~(\ref{Smatrix}) nor Eq.~(\ref{SmatrixTilde}) is
used for the analytic continuation. We do the continuation of the Jost matrix
that does not involve any additional square roots of the momenta.
Moreover, we use special semi-analytic representation of the Jost
matrices suggested in Refs.~\cite{our.MultiCh, our_Coulomb}, where the factors
responsible for the branching of the Riemann surface are given explicitly.
It was shown\cite{our_Coulomb} that the Jost matrices have the
following structure
\begin{eqnarray}
\label{multi.matrixelements}
   f^{(\mathrm{in/out})}_{mn}(E) &=&
   \frac{e^{\pi\eta_m/2}\ell_m!}{\Gamma(\ell_m+1\pm i\eta_m)}
   \left\{
   \frac{{C}_{\ell_n}(\eta_n)k_n^{\ell_n+1}}
   {{C}_{\ell_m}(\eta_m)k_m^{\ell_m+1}}{A}_{mn}(E)\ -\right.\\[3mm]
\nonumber
   &-&
   \left.\left[
   \frac{2\eta_mh(\eta_m)}{C_0^2(\eta_m)}\pm i\right]
   {C}_{\ell_m}(\eta_m){C}_{\ell_n}(\eta_n)
   k_m^{\ell_m}k_n^{\ell_n+1}{B}_{mn}(E)\right\}\ ,
\end{eqnarray}
where the matrices $\bm{A}(E)$ and $\bm{B}(E)$ are single-valued functions of
$E$, defined on a simple $E$-plane without any branch points. These matrices are
the same for both $\bm{f}^{\rm(in)}$ and $\bm{f}^{\rm(out)}$. It was also shown
that for real energies the matrices $\bm{A}(E)$ and $\bm{B}(E)$ are real.

Eq.~(\ref{multi.matrixelements}) gives us the Jost matrices in the
semi-analytic form, where the matrices $\bm{A}$ and $\bm{B}$ are unknown, but
are simple
and single-valued, while all the troublesome factors (that are responsible for
the branching) are given explicitly. They involve the Coulomb barrier
factor $C_\ell$ and the function $h(\eta)$ associated with the Coulomb
potential\footnote{The logarithmic branching stems from the function
$h(\eta)$.}:
\begin{equation}
\label{CL}
  C_\ell(\eta)=
  \frac{2^\ell e^{-\pi\eta/2}}{(2\ell)!!}
  \exp\left\{\frac12\left[\ln\Gamma(\ell+1+i\eta)+
  \ln\Gamma(\ell+1-i\eta)\right]\right\}
  \ \mathop{\longrightarrow}\limits_{\eta\to0}\ 1\ ,
\end{equation}
\begin{equation}
\label{h_function}
   h(\eta)=\frac12\left[\psi(i\eta)+
   \psi(-i\eta)\right]-\ln{\eta}\ ,
   \qquad
   \psi(z)=\frac{\Gamma'(z)}{\Gamma(z)}\ ,
   \qquad
   {\eta}=\frac{e^2Z_1Z_2\mu}{\hbar^2k}\ .
\end{equation}
For neutral particles (when $\eta=0$), Eq.~(\ref{multi.matrixelements}) becomes
more simple~\cite{our.MultiCh},
\begin{equation}
\label{multichannel.JostMatr.Jostfact}
   f_{mn}^{\rm(in/out)}(E)=
   \frac{k_n^{\ell_n+1}}{k_m^{\ell_m+1}}{A}_{mn}(E)\mp
   ik_m^{\ell_m}k_n^{\ell_n+1}{B}_{mn}(E)\ .
\end{equation}
It should be noted that the analytic structure of the $S$-matrix is much more
complicated than that of the Jost matrices. This becomes obvious if we
substitute the matrices (\ref{multi.matrixelements}) in Eq.~(\ref{Smatrix}). It
is therefore more simple to deal with $\bm{f}^{\rm(in/out)}(E)$, when doing the
analytic continuation from real to complex $E$. If the $S$-matrix itself is
used for such a task, and especially when such a matrix is constructed in a
simplified phenomenological way, it is difficult to guarantee that the
continuation is done to the correct sheet of the Riemann surface.

\subsection{Approximation and analytic continuation}
\label{sect.ApprCont}
The unknown matrices $\bm{A}(E)$ and $\bm{B}(E)$ in the semi-analytic
representations (\ref{multi.matrixelements}) and
(\ref{multichannel.JostMatr.Jostfact}) are single-valued and analytic. This
means that they can be expanded in Taylor series around an arbitrary complex
energy $E_0$,
\begin{equation}
\label{A.Taylor}
   \bm{A}(E)=\bm{a}^{(0)}+\bm{a}^{(1)}(E-E_0)+
   \bm{a}^{(2)}(E-E_0)^2+\cdots\ ,
\end{equation}
\begin{equation}
\label{B.Taylor}
   \bm{B}(E)=\bm{b}^{(0)}+\bm{b}^{(1)}(E-E_0)+
   \bm{b}^{(2)}(E-E_0)^2+\cdots\ ,
\end{equation}
where $\bm{a}^{(m)}(E_0)$ and $\bm{b}^{(m)}(E_0)$ are the $(N\times
N)$-matrices depending on the choice of the center $E_0$ of the expansion. These
matrices consist of unknown parameters. Taking the first several terms of these
expansions and finding the parameters $\bm{a}^{(m)}$ and $\bm{b}^{(m)}$ via
fitting some available experimental data, we can obtain approximate analytic
expressions (\ref{multi.matrixelements}) for the Jost matrices.

Since the matrices $\bm{A}(E)$ and $\bm{B}(E)$ are real for real energies, it is
convenient to choose $E_0$ on the real axis. As a result the parameters
$\bm{a}^{(m)}$ and $\bm{b}^{(m)}$ are also real. If they were complex then the
number of the fitting parameters would double (their real and imaginary
components are independent parameters).

Although $E_0$ is on the real axis, the expansions
(\ref{A.Taylor},\ref{B.Taylor}) are valid for complex $E$ within a circle around
$E_0$. After finding the fitting  parameters $\bm{a}^{(m)}$ and $\bm{b}^{(m)}$,
we can use the analytic expression (\ref{multi.matrixelements}) for the Jost
matrix $\bm{f}^{\mathrm{(in)}}(E)$, to locate the resonances as the roots of
Eq.~(\ref{detfzero}) at complex energies. When doing this, we can choose the
appropriate sheet of the Riemann surface. The single-valued functions
$\bm{A}(E)$ and $\bm{B}(E)$ are the same on all the sheets. The differencies
only stem from the explicit factors depending on $k_n$ in
Eq.~(\ref{multi.matrixelements}).

When the energy $E$ is given, in order to choose a specific sheet, we calculate
the square roots (\ref{ch_momenta}) for all the channel
momenta and check the signs of their imaginary parts. If for a particular $k_n$
the sign of $\mathrm{Im}\,k_n$ is not right, we simply replace $k_n$ with
$-k_n$. After getting all the channel momenta with the appropriate signs, we
calculate the Jost matrices as is given by Eq.~(\ref{multi.matrixelements}).
Since in this equation all the factors depending on the choice of the signs,
are given in an exact way, we are sure that despite the approximations
(\ref{A.Taylor},\ref{B.Taylor}), the Jost matrices are calculated just for a
chosen sheet of the Riemann surface. In other words, the analytic continuation
of the Jost matrices from the real axis (where the fitting is done) to a chosen
Riemann sheet is always done correctly.

Loking at the expansions (\ref{A.Taylor},\ref{B.Taylor}) , which we truncate at
certain number of terms, one might think that we work with a polynomial
approximation of the reaction cross section. However, since the $S$-matrix is a
``ratio'' (\ref{Smatrix}) of the Jost matrices, our approximation is more
similar to the Pad\'{e} approximation~\cite{Press}. It is well known that the
Pad\'{e} approximants, which are ratios of two polynomials, are more accurate
and efficient than any polynomial ones, because they include some complex poles
that influence the behaviour of the function on the real axis. In our approach,
each element of the $S$-matrix is a ratio of two functions. They are not just
polynomials but some more complicated functions. For all the elements, the
denominator is the same, namely, $\det \bm{f}^{\mathrm{(in)}}(E)$. Our
approximation is better than the  Pad\'{e} one because it  involves not only the
poles but also the correct factors determining the branching of the Riemann
surface.

\section{The data to fit}
The unstable system ${}^5\mathrm{He}$ has been studied for many years. There is
an extensive  list of publications~\cite{Tilley} where  the results of these
studies are reported. In principle, using the approach described in the present
paper it is possible to fit any available data. Indeed, the parametrized Jost
matrices give us the corresponding $S$-matrix, from which it is possible to
calculate any observable quantity as a function of the fitting parameters.
However, it would be a waste of time and effort to collect and fit the original
data. Instead, we can rely on the existing $R$-matrix fits. In fact the
$R$-matrix fit is the most accurate and efficient way of parametrizing a
collection of scattering data.

The $R$-matrix is good for real energies, but its extension to complex $E$ is
questionable. This should be clear from the previous Sections of this paper,
where we describe the complicated topology of the Riemann surface. The analytic
continuation onto this surface is the task where our approach can complement the
real energy $R$-matrix parametrization.

Taking an $R$-matrix with a given set of parameters, we calculate the
partial cross sections (in certain channels) at the energies we need. Thus
obtained cross sections can be considered
as experimental data, which we then fit using the Jost matrices. In a sense,
within such an approach the $R$-matrix fit serves as a ``preliminary
refinement'' of the ``raw'' data. It is similar to doing the phase-shift
analysis. Nobody actually measures the scattering phase shifts, but still the
tables of them are considered as experimental data.

As was stated from the outset, we restrict our consideration to the resonant
state of ${}^5\mathrm{He}$ with $J^\pi=\frac32^+$ near the $dt$-threshold. This
state can only be a mixture of the four channels listed in
Table~\ref{table.channels}, i.e. its wave function is a linear combination of
the four individual channel wave functions. Apparently, their contributions into
such a combination are different. In order to do a preliminary estimate of the
importance of each of the channels in this combination, we can look at the
partial cross sections defined by Eq.~(\ref{particular_sigma}), for all
(sixteen) possible mutual transitions among the four channels. This can be done
using an $R$-matrix parametrization.

\begin{figure}
\centerline{\epsfig{file=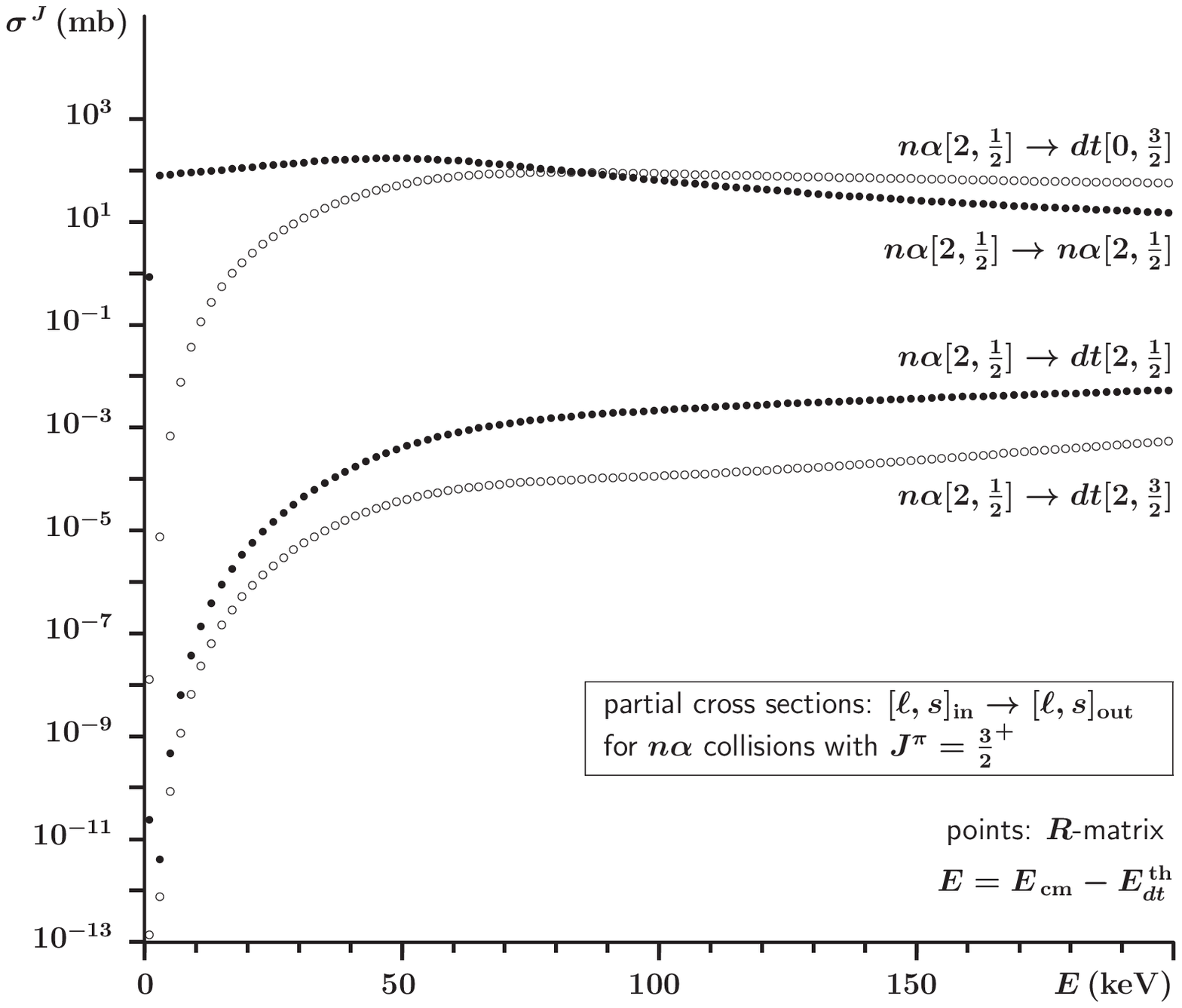}}
\caption{\sf
Partial cross sections for the channels $1\to1$, $1\to2$, $1\to3$, and $1\to4$
as is denoted in Table~\ref{table.channels}.
}
\label{fig.log1}
\end{figure}

\begin{figure}
\centerline{\epsfig{file=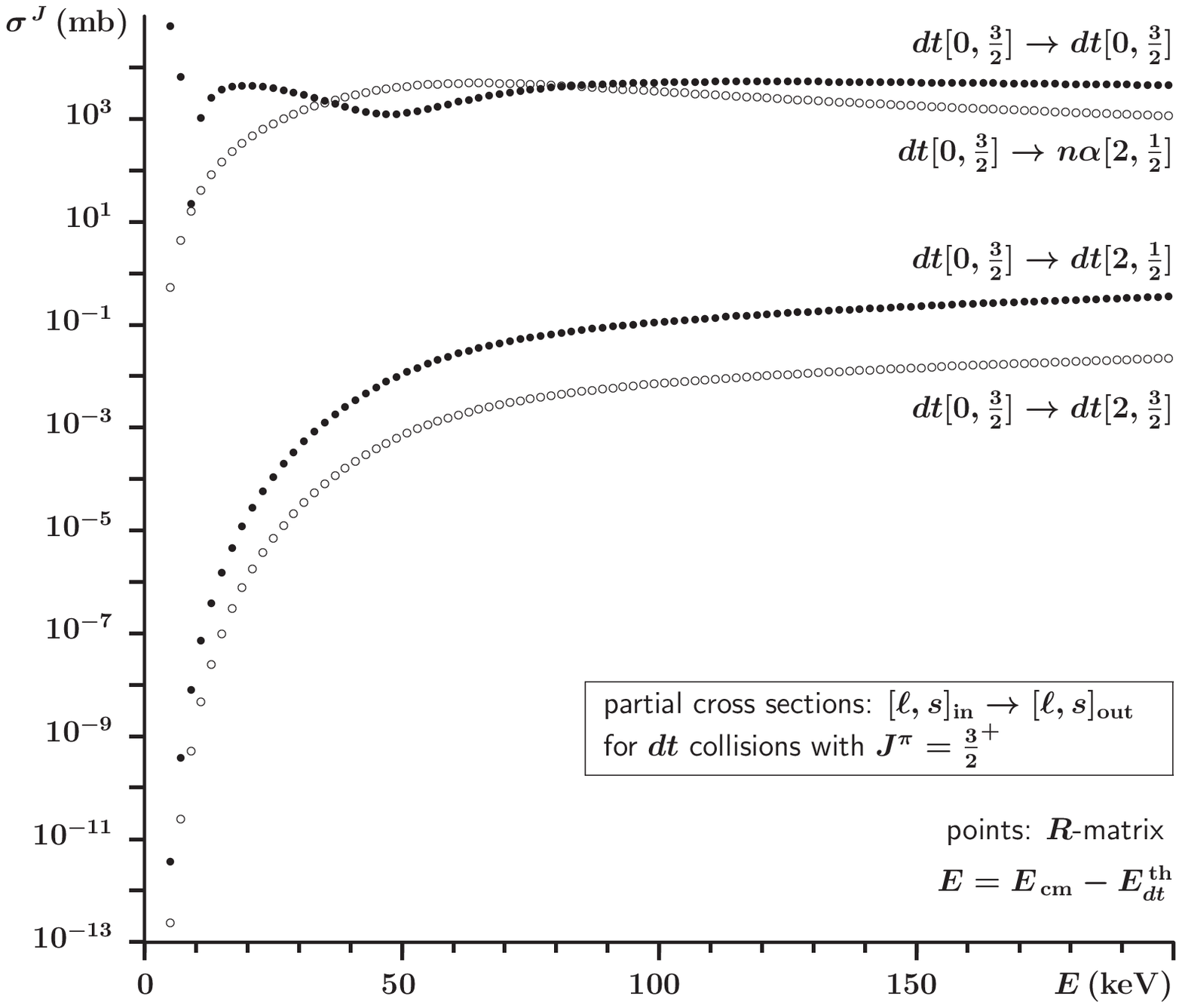}}
\caption{\sf
Partial cross sections for the channels $2\to1$, $2\to2$, $2\to3$, and $2\to4$
as is denoted in Table~\ref{table.channels}.
}
\label{fig.log2}
\end{figure}

\begin{figure}
\centerline{\epsfig{file=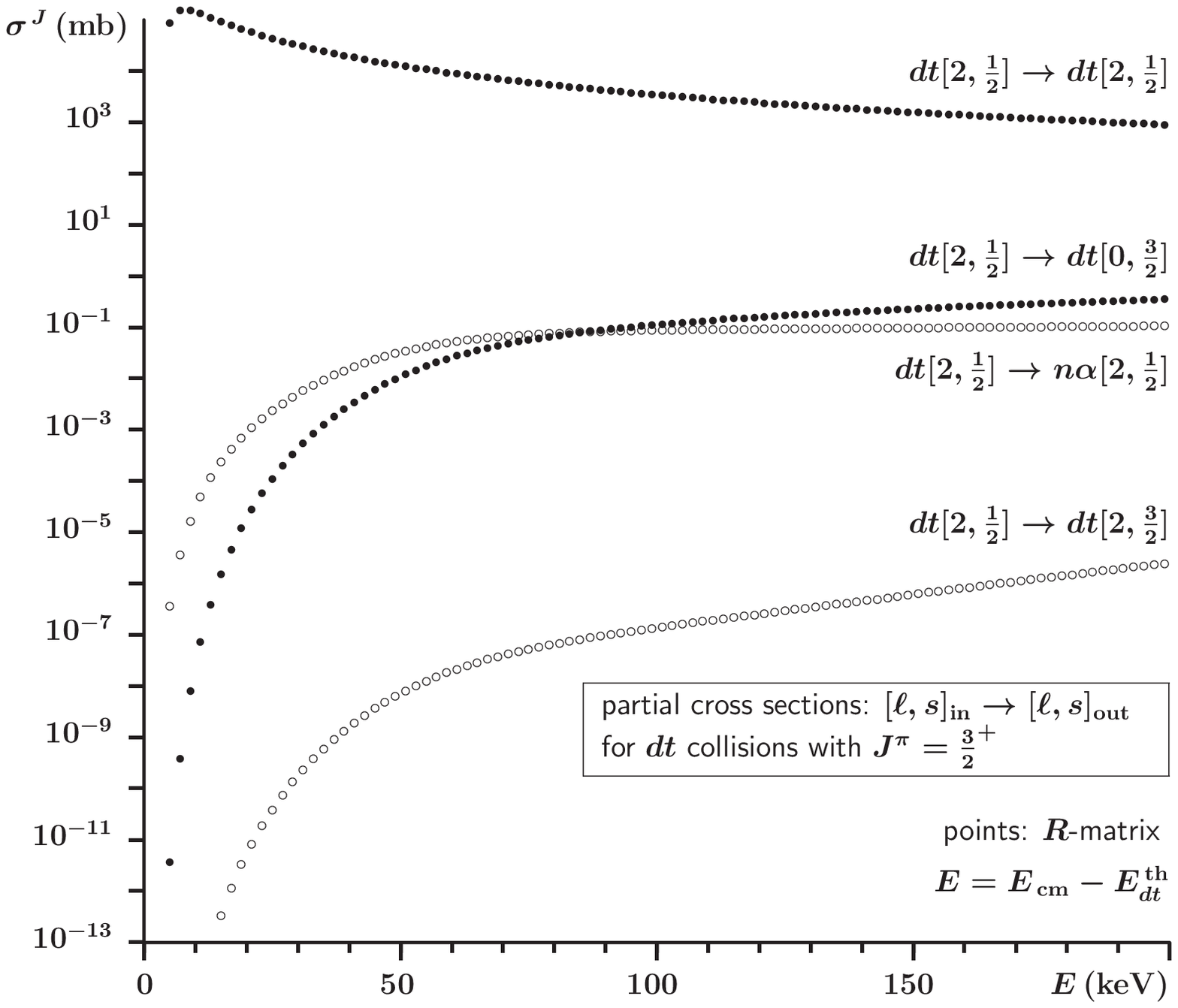}}
\caption{\sf
Partial cross sections for the channels $3\to1$, $3\to2$, $3\to3$, and $3\to4$
as is denoted in Table~\ref{table.channels}.
}
\label{fig.log3}
\end{figure}

\begin{figure}
\centerline{\epsfig{file=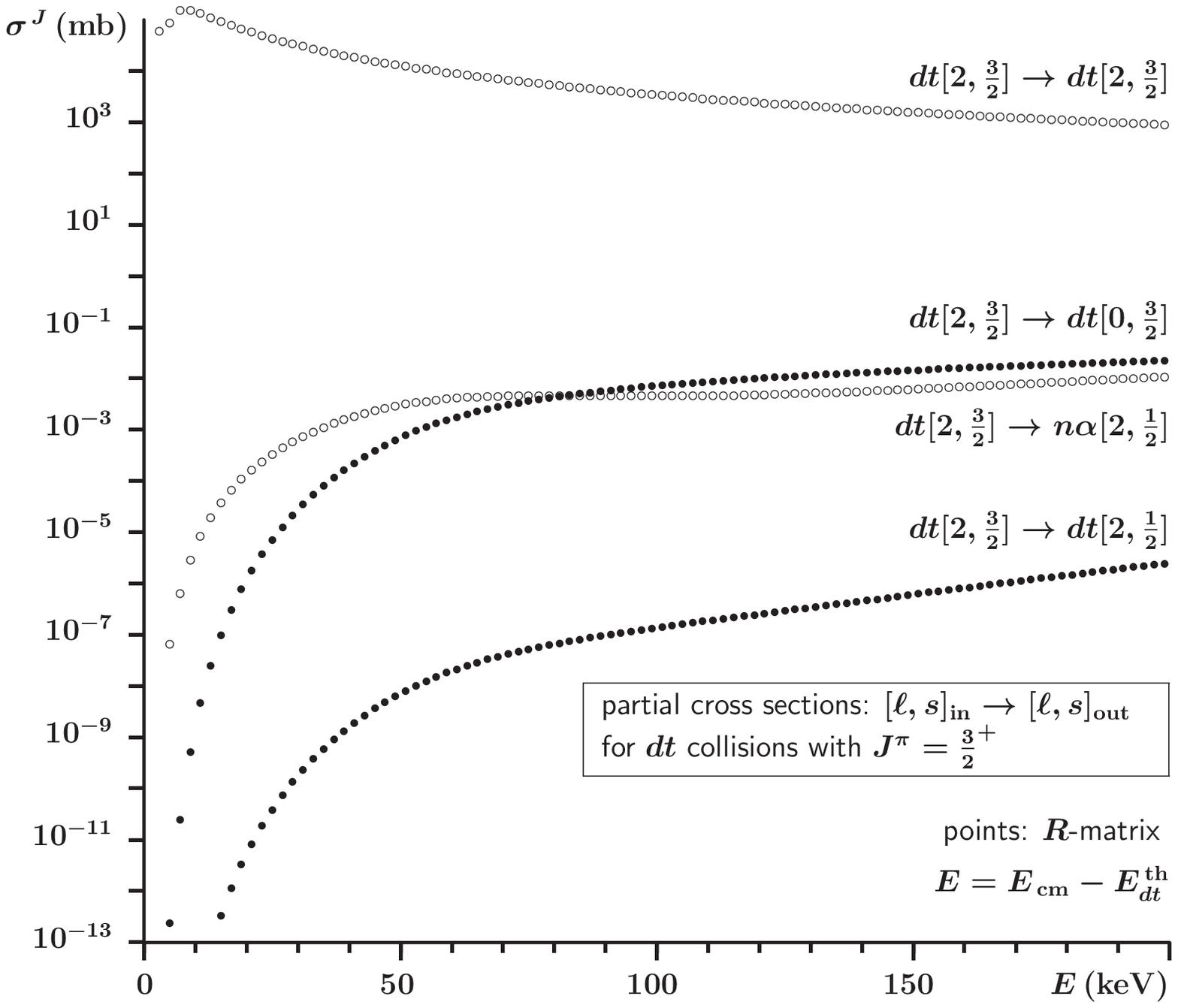}}
\caption{\sf
Partial cross sections for the channels $4\to1$, $4\to2$, $4\to3$, and $4\to4$
as is denoted in Table~\ref{table.channels}.
}
\label{fig.log4}
\end{figure}

As such a parametrization, we use the comprehensive four-level four-channel
$R$-matrix fit reported in Ref.~\cite{PRL59}. Generated from this $R$-matrix,
the partial cross sections for the transitions from the first, second, third,
and the fourth channels, respectively, are shown in
Figs.~\ref{fig.log1}-\ref{fig.log4}.  The magnitudes of the cross
sections for the transitions to different channels are so different that we
have to use the logarithmic scale in order to fit them in the same figure.

As is seen, at the energies around the resonance we are studying
($\sim50$\,keV above the $dt$-threshold), the probabilities of the off-diagonal
transitions
$1\leftrightarrow3$, $1\leftrightarrow4$, $2\leftrightarrow3$,
$2\leftrightarrow4$, and $3\leftrightarrow4$ are at least five orders of
magnitude smaller than those for the diagonal ones. They also are several
orders of magnitude smaller than the transition
probabilities between the first and the second channels. Coming back to
Table~\ref{table.channels}, it is not difficult to qualitatively explain some of
these differences.

Indeed, in the first channel
there is only the centrifugal barrier with $\ell=2$, which is not an obstacle
at the $n\alpha$-collision energy $\sim18$\,MeV. The neutron and the
$\alpha$-particle
can easily come close to each other and form the compound resonant state. When
this state decays, the fragments $n$ and $\alpha$ can easily escape through
that centrifugal barrier.

The second channel ($dt$ with $\ell=0$) has only the Coulomb barrier. The
deuteron and triton still can penetrate through it to form the resonance and
then to escape from it. However the relative $dt$-energy is much lower than in
the first channel and thus the channel 2 should have a weaker coupling with the
resonance. In particular, the partial decay width into this channel should be
smaller than the width for the $n\alpha$-channel (we will see a quantitative
confirmation of this intuitive reasoning in Sec.~\ref{sec.results}).

The channels 3 and 4 have the sum of both the centrifugal and the Coulomb
barriers. When the $dt$-system is in one of these channels and the energy is
around $\sim50$\,keV, the penetration probability for the total barrier is too
small. This implies that the couplings of these channels with the resonance are
extremely weak.

The channels can couple to each other either via the compound resonance or via
the direct reactions of the pick-up and stripping of the proton. As we see, the
resonance can only couple the channels 1 and 2. In a direct reaction the
transferred proton has to tunnel through the same barriers, which means that for
the channels 3 and 4 such a process is also extremely improbable. The same
conclusion can be found, for example, in Ref.~\cite{Melezhik}.

In Figs. \ref{fig.log3} and \ref{fig.log4} it is seen that the transitions
$3\leftrightarrow4$ have the cross sections $\sim10^{-7}$\,mb (the smallest of
all the inter-channel transitions). This cannot be explained as in the above,
because the barriers for these channels are the same. We do not have any
reasonable explanation why the spin state 1/2 does not ``want'' to transit to
the state 3/2 and vise versa. But this is the result of the $R$-matrix fit of
the experimental data. We therefore simply accept as a fact that the channels 3
and 4 are practically not coupled to each other.

Summarizing the above reasoning, we conclude that the channels 1 and 2 are
strongly coupled with each other, while the channels 3 and 4 are not coupled to
any other channels and can be treated individually,
\begin{equation}
\label{couplings12}
\begin{array}{lcl}
   n\alpha\ (2,\frac12) &  & n\alpha\ (2,\frac12)\\
   & \displaystyle\genfrac{}{}{0pt}{}{\searrow}{\nearrow}
   \bigcirc
     \displaystyle\genfrac{}{}{0pt}{}{\nearrow}{\searrow}
   & \\
   dt\ (0,\frac32) &  & dt\ (0,\frac32)\\
\end{array}\ ,
\end{equation}
\begin{eqnarray}
\label{couplings3}
   dt\ (2,\textstyle\frac12) & \longrightarrow & dt\ (2,\textstyle\frac12)\ ,
\end{eqnarray}
\begin{eqnarray}
\label{couplings4}
   dt\ (2,\textstyle\frac32) & \longrightarrow & dt\ (2,\textstyle\frac32)\ .
\end{eqnarray}
This means that our original four-channel problem (as is given in
Table~\ref{table.channels}) is divided in a two-channel problem (coupled
channels 1 and 2) and two single-channel problems (channels 3 and 4). Of course
this is an approximation, but the extremely small
inter-channel cross sections connecting the channels (1,2) with 3 and 4, allow
us to ignore them. These cross sections are at least five orders of magnitude
smaller than those we take into account. This is beyound the accuracy of our
calculations.

\section{Fitting procedure}
\label{sec.procedure}
The method for fitting the experimental data with the parametrized Jost
matrices was tested on several single-channel and two-channel
model-problems~\cite{our.exp.shortrange,our.exp.Coulomb}, where the exact
locations of resonances were known. It was demonstrated that the method is
stable and reliable. It works equally well in the problems for neutral and for
charged particles. The procedure we use here is basically the same.

The matrices $\bm{A}(E)$ and $\bm{B}(E)$ in Eq.~(\ref{multi.matrixelements}),
which are analytic and single-valued functions of the energy, are approximated
by the $(M+1)$ Taylor terms,
\begin{eqnarray}
\label{Aapprox}
   A_{n'n}(E) &\approx& \sum_{m=0}^M a^{(m)}_{n'n}(E-E_0)^m\ ,\\[3mm]
\label{Bapprox}
   B_{n'n}(E) &\approx& \sum_{m=0}^M b^{(m)}_{n'n}(E-E_0)^m\ ,
   \qquad n',n=1,2,\dots,N\ ,
\end{eqnarray}
where $a^{(m)}_{n'n}$ and $b^{(m)}_{n'n}$ are the fitting parameters and $N=2$
is the number of the coupled channels. The Jost matrices
(\ref{multi.matrixelements}) with these $\bm{A}$ and $\bm{B}$ are then used to
calculate the $S$-matrix (\ref{Smatrix}) and the approximate partial cross
sections (\ref{particular_sigma}), $\tilde{\sigma}_{n'\gets n}$, which thus
depend on the fitting parameters.

In order to find the optimal values of the fitting parameters, we minimize the
following $\chi^2$-function
\begin{equation}
\label{chisquare}
   \chi^2=\sum_{n',n=1}^N\sum_{i=1}^K\left|
    \tilde{\sigma}_{n'\gets n}(E_i)-\sigma_{n'\gets n}(E_i)\right|^2\ ,
\end{equation}
where $K$ is the number of experimental points taken into
account for each partial cross section, and $\sigma_{n'\gets n}(E_i)$ is the
corresponding experimental cross section at the energy $E_i$. Since the
experimental data are taken from the $R$-matrix analysis, the experimental
errors are not defined. We therefore put all of them to unity in the
$\chi^2$-function (\ref{chisquare}).

The minimization is done using the MINUIT program developed in
CERN~\cite{MINUIT}. The function (\ref{chisquare}) has many local minima. In
order to find the best of them, we repeat the procedure many times ($\sim1000$)
with randomly chosen starting values of the parameters. After a good minimum
is found, we refine it by choosing random starting point around the best point
found in the parameter space. After each improvement, we do the random choice
of the starting parameters around the new best point.

It should be noted that the observables are expressed via the elements of the
$S$-matrix (\ref{Smatrix}), which is the ratio of the Jost matrices. This means
that any common factor in $\bm{f}^{(\mathrm{in})}$ and
$\bm{f}^{(\mathrm{out})}$ cancels
out in the $S$-matrix. Therefore the set of the parameters $a^{(m)}_{n'n}$ and
$b^{(m)}_{n'n}$ can be scaled by any convenient factor. This does not affect
any results. For example, one of the parameters can be put to a fixed value,
which reduces their number in one parameter, from $2(M + 1)N^2$ to
$2(M + 1)N^2-1$.

\section{Results}
\label{sec.results}
\begin{figure}
\centerline{\epsfig{file=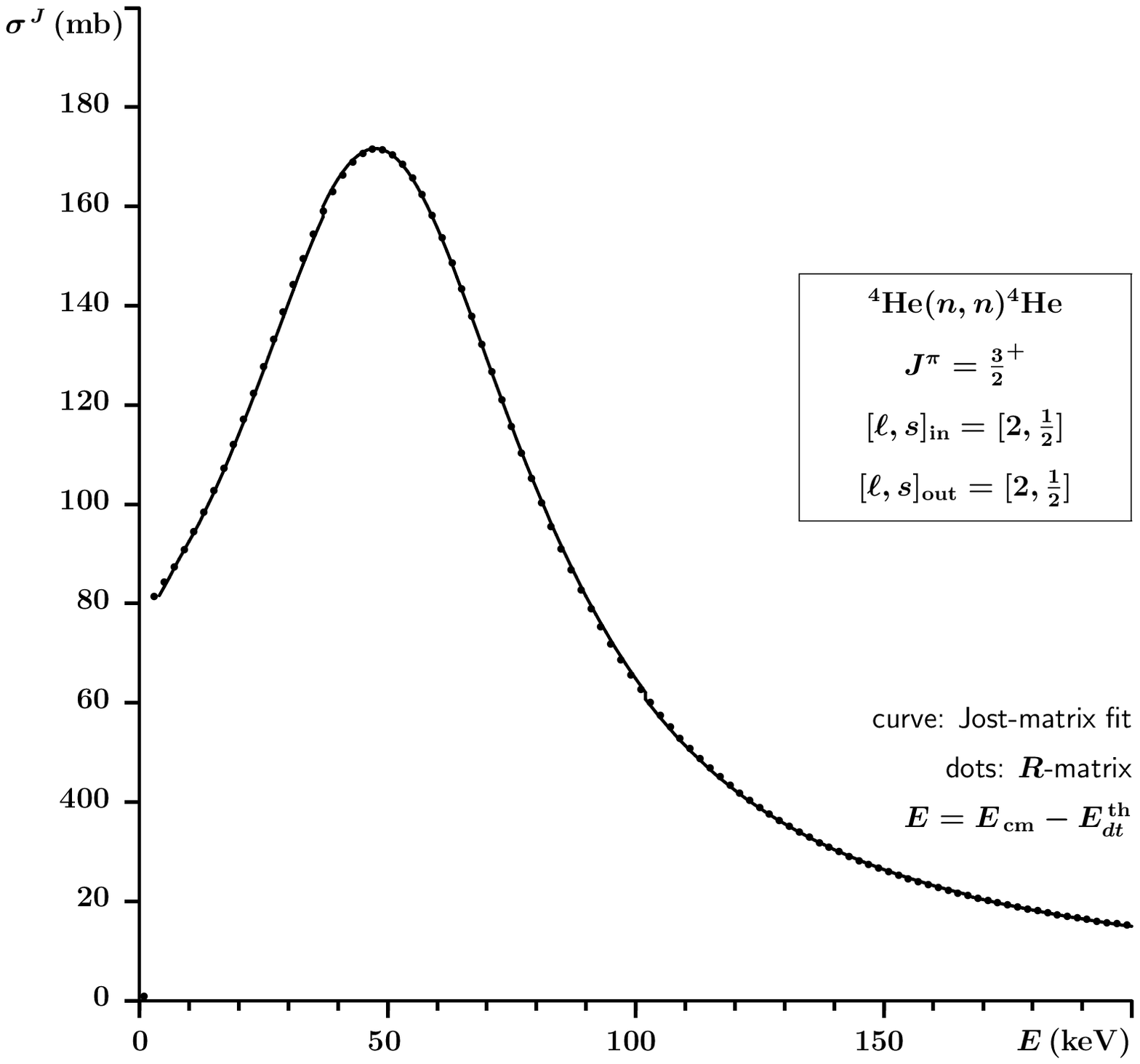}}
\caption{\sf
Fit of the partial cross section for the elastic transition $1\to1$
(see the notation in Table~\ref{table.channels}).
}
\label{fig.fit11}
\end{figure}

\begin{figure}
\centerline{\epsfig{file=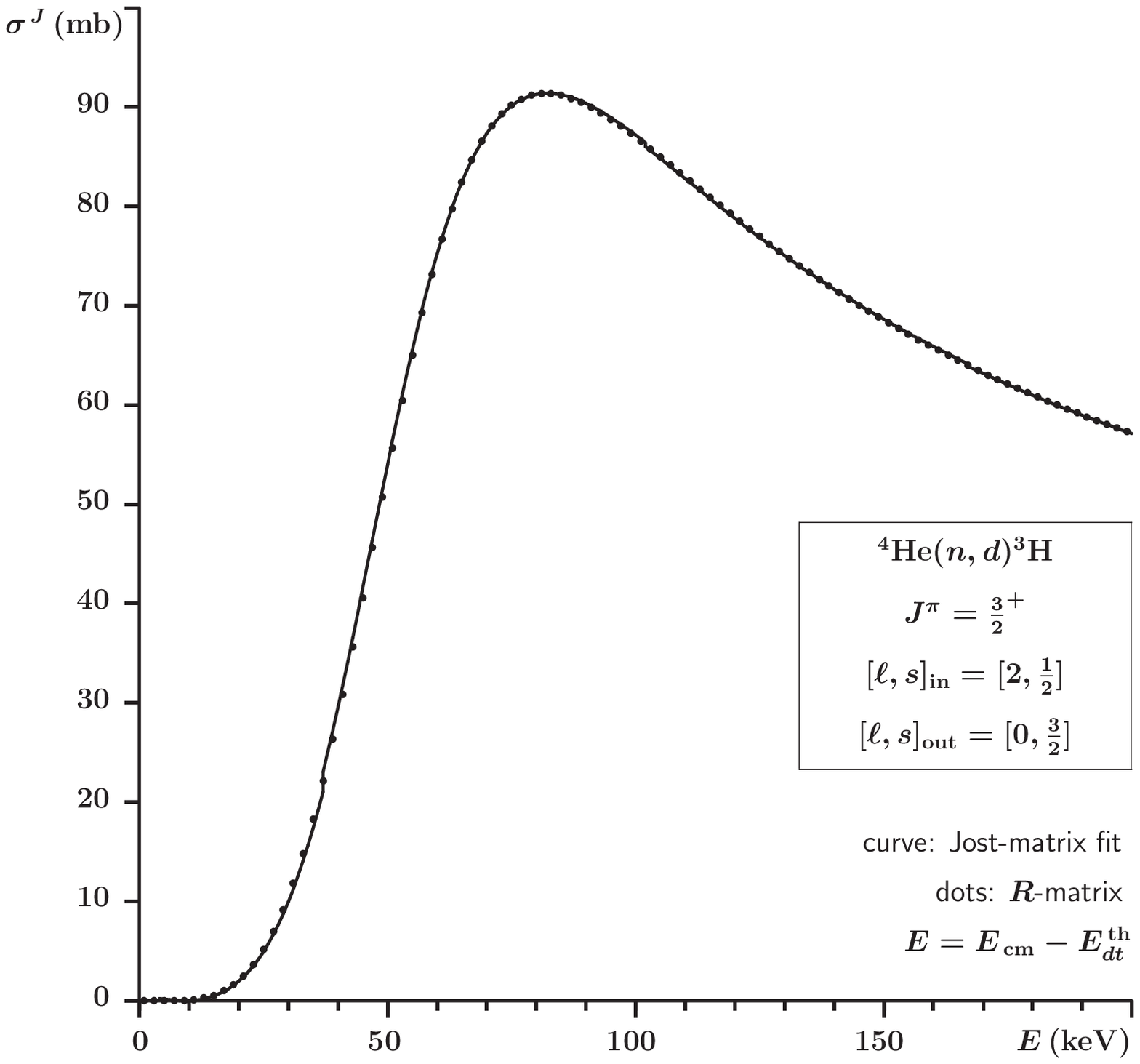}}
\caption{\sf
Fit of the partial cross section for the inelastic transition $1\to2$
(see the notation in Table~\ref{table.channels}).
}
\label{fig.fit12}
\end{figure}

\begin{figure}
\centerline{\epsfig{file=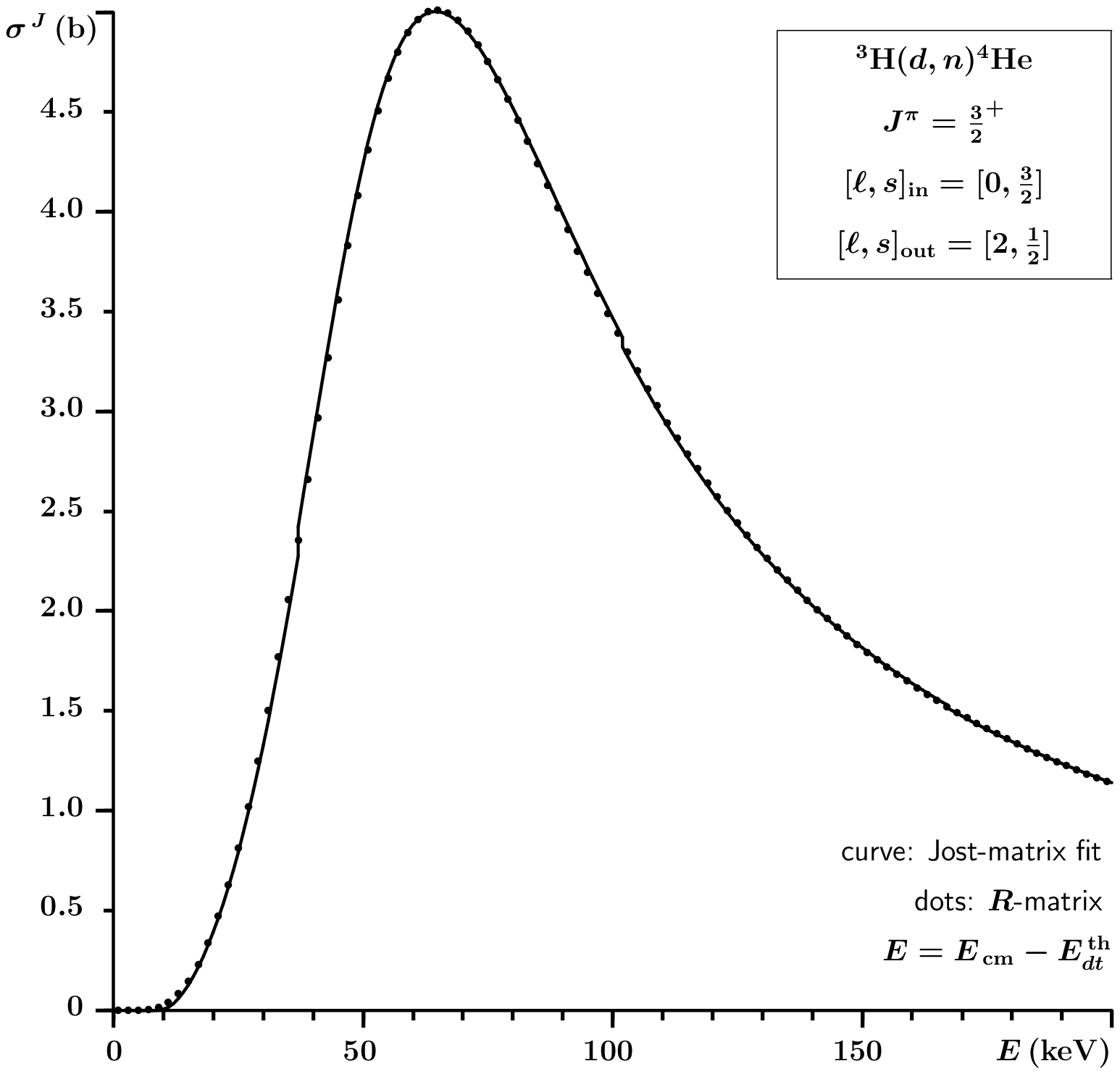}}
\caption{\sf
Fit of the partial cross section for the inelastic transition $2\to1$
(see the notation in Table~\ref{table.channels}).
}
\label{fig.fit21}
\end{figure}

\begin{figure}
\centerline{\epsfig{file=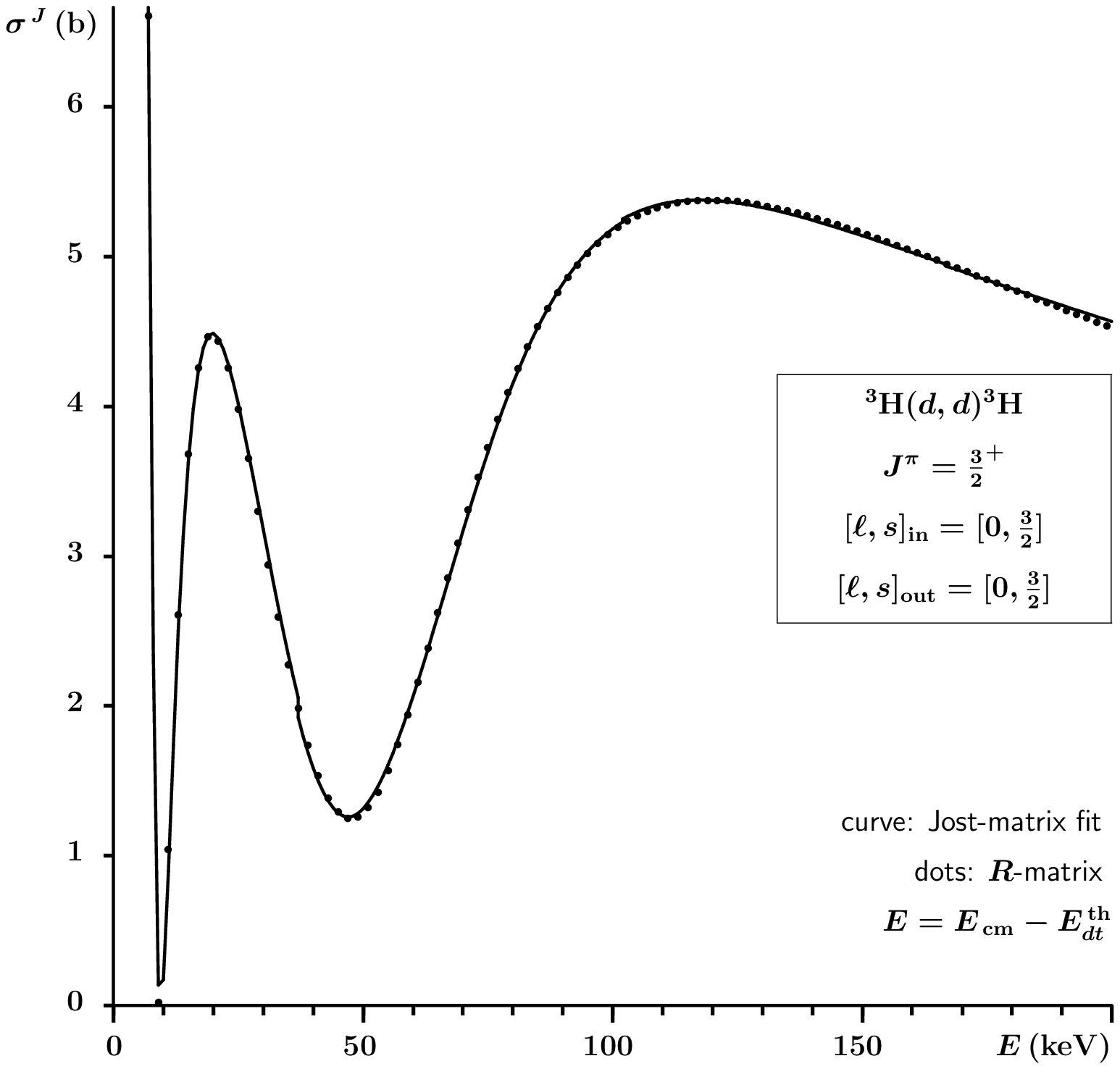}}
\caption{\sf
Fit of the partial cross section for the elastic transition $2\to2$
(see the notation in Table~\ref{table.channels}).
}
\label{fig.fit22}
\end{figure}

The two-channel ($N=2$) data (obtained from the $R$-matrix given in
Ref.~\cite{PRL59}) for the four processes (\ref{couplings12}) were fitted as is
decribed in Sec.~\ref{sec.procedure} with $M=4$ and $E_0=50$\,keV (the
energy is counted from the $dt$-threshold). The results of the fit are
graphically shown in Figs. \ref{fig.fit11}, \ref{fig.fit12}, \ref{fig.fit21},
and \ref{fig.fit22}.

This fit was achieved with the set of parameters given in Table
\ref{table.parameters12}. The units for the parameters are chosen in such a way
that the Jost matrices are dimensionless.

\begin{table}
\begin{center}
\begin{tabular}{|c|c|c|c|c|c|}
\hline
\multicolumn{3}{|c|}{$E_0$}
& $40\,\mathrm{keV}$ & $50\,\mathrm{keV}$ & $60\,\mathrm{keV}$\\
\hline
$m$ & $n'$ & $n$
&
\parbox{2cm}{\begin{center}
$a_{n'n}^{(m)}$,\ $b_{n'n}^{(m)}$\\[1mm]
$[\mathrm{MeV}^{-m}]$\end{center}}
&
\parbox{2cm}{\begin{center}
$a_{n'n}^{(m)}$,\ $b_{n'n}^{(m)}$\\[1mm]
$[\mathrm{MeV}^{-m}]$\end{center}}
&
\parbox{2cm}{\begin{center}
$a_{n'n}^{(m)}$,\ $b_{n'n}^{(m)}$\\[1mm]
$[\mathrm{MeV}^{-m}]$\end{center}}\\
\hline
0 & 1 & 1 &
$         0.11653    $,\   $          53.625    $ & 
$0.34093$,\  $55.416$ & 
$          3.3788    $,\   $          48.166    $\\ 
  & 1 & 2 &
$          1.0813    $,\   $         -2.1316    $ & 
$0.77509$,\  $-2.2752$ &
$         0.57354    $,\   $         -2.0536    $\\
  & 2 & 1 &
$         0.96416    $,\   $          32.904    $ & 
$0.60767$,\  $-3.9457$ &
$         0.35755    $,\   $         -24.660    $\\
  & 2 & 2 &
$         0.057233$,\   $          4.5797    $ & 
$0.063612$,\  $4.5766$ &
$         0.071082$,\   $          4.4843    $\\
\hline
1 & 1 & 1 &
$          512.04    $,\   $         -221.91    $ & 
$537.08$,\  $148.21$ &
$          656.88    $,\   $          410.14    $\\
  & 1 & 2 &
$         -17.408    $,\   $         -4.4977    $ & 
$-6.0448$,\  $-16.604$ &
$          5.7879    $,\   $         -26.110    $\\
  & 2 & 1 &
$          13.264    $,\   $         -1127.4    $ & 
$8.0623$,\  $-1102.8$ &
$          20.584    $,\   $         -1038.6    $\\
  & 2 & 2 &
$          2.4459    $,\   $          96.388    $ & 
$2.7856$,\  $115.63$ &
$          3.1034    $,\   $          124.21    $\\
\hline
2 & 1 & 1 &
$         -6457.5    $,\   $          14179    $ & 
$10660$,\  $15358$ &
$          20237    $,\   $          12090    $\\
  & 1 & 2 &
$          567.39    $,\   $         -705.56    $ & 
$733.01$,\  $-780.22$ &
$          678.85    $,\   $         -696.52    $\\
  & 2 & 1 &
$          54.500    $,\   $          30091    $ & 
$967.39$,\  $11746$ &
$          1699.3    $,\   $          2380.9    $\\
  & 2 & 2 &
$          36.874    $,\   $          1314.1    $ & 
$57.122$,\  $1841.8$ &
$          66.282    $,\   $          2116.1    $\\
\hline
3 & 1 & 1 &
$         737310$,\   $          29257    $ & 
$659610$,\  $-31260$ &
$         506310$,\   $         -58864    $\\
  & 1 & 2 &
$          13254    $,\   $         -2728.6    $ & 
$8435.9$,\  $-3067.5$ &
$          6439.0    $,\   $         -2442.2    $\\
  & 2 & 1 &
$          34325    $,\   $        -352760$ & 
$47376$,\  $-94790$ &
$          45900    $,\   $          95653    $\\
  & 2 & 2 &
$          1082.3    $,\   $          34639    $ & 
$1148.7$,\  $33655$ &
$          1036.4    $,\   $          26021    $\\
\hline
4 & 1 & 1 &
$         3123300$,\   $        -2214800$ & 
$2140300$,\  $-1764300$ &
$         1353200$,\   $        -1201600$\\
  & 1 & 2 &
$         119530$,\   $         -37503    $ & 
$75737$,\  $-14863$ &
$          50499    $,\   $         -5779.9    $\\
  & 2 & 1 &
$         774390$,\   $         11616000$ & 
$565470$,\  $9340700$ &
$         373110$,\   $         6407400$\\
  & 2 & 2 &
$          22306    $,\   $         171080$ & 
$12689$,\  $56199$ &
$          7858.0    $,\   $          14180    $\\
\hline
\end{tabular}
\end{center}
\caption{\sf
Parameters of the expansions (\ref{Aapprox},\ref{Bapprox}) for the coupled
channels 1 and 2 with three choices of the central point $E_0$. These
parameters  for $E_0=50$\,keV were used to generate the curves shown in Figs.
\ref{fig.fit11}, \ref{fig.fit12}, \ref{fig.fit21}, and \ref{fig.fit22}, and to
locate the resonances.
}
\label{table.parameters12}
\end{table}

As is seen, in
Eq.~(\ref{multi.matrixelements}) different matrix elements of the Jost matrices
may involve different powers of the channel momenta (in the explicit factors).
As a result, the units of $A_{n'n}$ or $B_{n'n}$ for different subscripts are
different as well. To avoid this,
we can always replace the channel momenta $k_n$ in this equation with the
product $k_nR$ with an arbitrarily chosen radius $R$. Since the matrices
$\bm{A}$ and $\bm{B}$ are unknown, the thus appearing additional constant
factors (powers of $R$) can be included in $\bm{A}$ and $\bm{B}$. Such an
inclusion cannot change their analytic properties, namely, the fact that they
are single-valued functions of the energy. For us it is convenient to choose
$R=1$\,fm.

\begin{figure}
\centerline{\epsfig{file=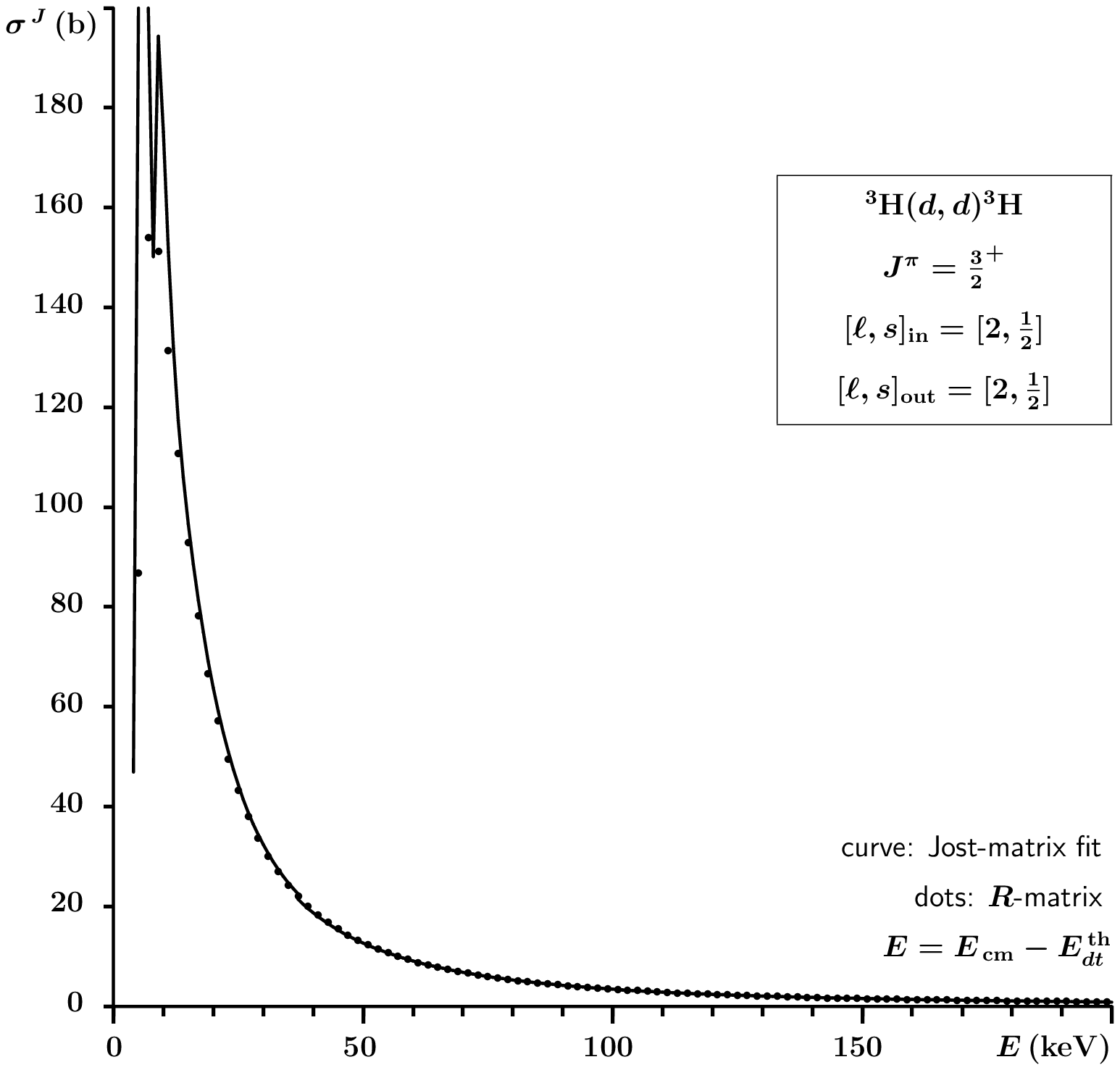}}
\caption{\sf
Fit of the partial cross section for the elastic transition $3\to3$
(see the notation in Table~\ref{table.channels}).
}
\label{fig.fit33}
\end{figure}

\begin{figure}
\centerline{\epsfig{file=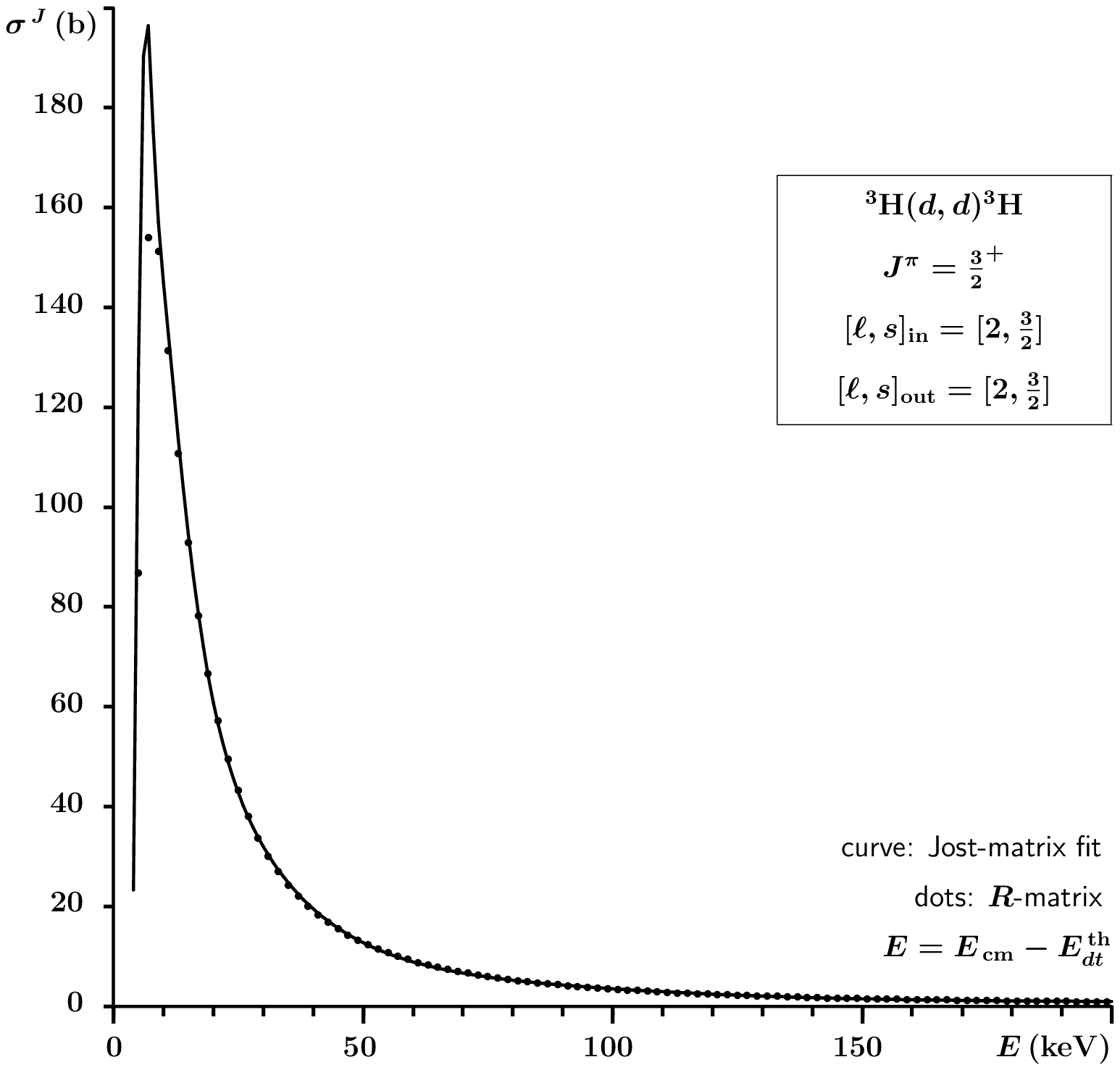}}
\caption{\sf
Fit of the partial cross section for the elastic transition $4\to4$
(see the notation in Table~\ref{table.channels}).
}
\label{fig.fit44}
\end{figure}

The single-channel data (obtained from the $R$-matrix given in
Ref.~\cite{PRL59}) for the processes (\ref{couplings3}) and (\ref{couplings4})
were fitted in the same way with $M=4$ and $E_0=50$\,keV. The results of the
fit are graphically shown in Figs. \ref{fig.fit33} and \ref{fig.fit44}, and the
corresponding parameters are given in Table~\ref{table.parameters34}.
The much larger values of these $D$-wave cross sections as compared to the 
$S$-wave ones shown in Fig.~\ref{fig.fit22}, may cause a confusion.
It is therefore worthwhile to add some explanations concerning these elastic 
cross sections. 

\begin{table}
\begin{center}
\begin{tabular}{|c|c|c|}
\hline
    & $3\to3$ & $4\to4$\\
\cline{2-3}
$m$ &
$a^{(m)}$,\ $b^{(m)}$\ $[\mathrm{keV}^{-m}]$
&
$a^{(m)}$,\ $b^{(m)}$\ $[\mathrm{keV}^{-m}]$\\
\hline
0 & $        -2.9864\times10^{-2}$,\   $        -154870$ &
$        -2.3603\times10^{-2}$,\   $        -118470$\\
\hline
1 & $         -1.4516\times10^{-3}    $,\   $        -413.64$ &
$         -1.2782\times10^{-3}    $,\   $        -872.71$\\
\hline
2 & $         -2.0099\times10^{-5}    $,\   $        -2.1701$ &
$         -2.5236\times10^{-5}    $,\   $        -17.221$\\
\hline
3 & $         -4.2416\times10^{-9}    $,\   $         2.0267\times10^{-2}$ &
$         -2.2213\times10^{-7}    $,\   $         9.9099\times10^{-2}$\\
\hline
4 & $          1.1253\times10^{-9}    $,\   $  -4.5799\times10^{-5}$ &
$         -7.9892\times10^{-10}    $,\   $  -1.3967\times10^{-3}$\\
\hline
\end{tabular}
\end{center}
\caption{\sf
Parameters of the expansions (\ref{Aapprox},\ref{Bapprox}) with
$E_0=50\,\mathrm{keV}$ for the (uncoupled) single
channels 3 and 4. These parameters generate the curves shown in Figs.
\ref{fig.fit33} and \ref{fig.fit44}.
}
\label{table.parameters34}
\end{table}

They are the partial cross sections integrated over all the scattering angles. 
Usually, the differential cross section for the charged particles is given as 
the absolute square of the sum of the pure Coulomb and the nuclear amplitudes.
The Coulomb amplitude is infinite in the forward direction. The integrated 
cross section is therefore infinite as well. As a consequence, the partial wave
decomposition of the pure Coulomb amplitude does not converge. However, each 
partial Coulomb cross section remains finite but increases with $\ell$. In our 
case, the nuclear part of the partial-wave scattering amplitude is known from 
the $R$-matrix fit. By adding to it the corresponding partial-wave Coulomb 
amplitude and integrating over the angles, we obtain the ``experimental'' 
partial cross section (the dots in the Figures). We have to add the pure 
Coulomb part to the $R$-matrix amplitude because, these ``experimental'' data 
are then fitted using the Jost matrices where the Coulomb and nuclear 
contributions are not separated. The ``experimental'' points in 
Figs.~\ref{fig.fit33} and \ref{fig.fit44} are higher than the points in 
Fig.~\ref{fig.fit22} because the pure Coulomb contribution grows with the 
angular momentum.

When looking for resonances, we considered the Jost matrix
(\ref{multi.matrixelements}) at complex energies. The zeros of its determinant
were sought using the Newton's method~\cite{Press}.
The choice of an appropriate sheet of the Riemann surface was
done as is described in Sec.~\ref{sect.ApprCont}.

The partial widths for the two-channel system were determined using the method
developed in Ref.~\cite{my.partial}. Actually, they are expressed in terms of
the matrix elements of the Jost matrices:
\begin{equation}
\label{Gpartial}
   \Gamma_i=
   \displaystyle\frac{\mathrm{Re}(k_i)|\mathcal{A}_i|^2\Gamma}
   {\displaystyle\sum_{i'=1}^N\frac{\mu_i}{\mu_{i'}}
   \mathrm{Re}(k_{i'})|\mathcal{A}_{i'}|^2}\ ,
\end{equation}
where $i=1$ and $i=2$ correspond to the $n\alpha$ and $dt$ channels, and
$\mathcal{A}_1$ and $\mathcal{A}_2$ are  the asymptotic amplitudes
(see Ref.~\cite{my.partial}) of the channels, given by
\begin{equation}
\label{A_1A_2_Final}
   \mathcal{A}_1=f^{(\mathrm{out})}_{11}-\displaystyle
   \frac{f^{(\mathrm{in})}_{11}f^{(\mathrm{out})}_{12}}{f^{(\mathrm{in})}_{12}}
   \ ,\qquad
   \mathcal{A}_2=f^{(\mathrm{out})}_{21}-\displaystyle
   \frac{f^{(\mathrm{in})}_{11}f^{(\mathrm{out})}_{22}}{f^{(\mathrm{in})}_{12}}
   \ .
\end{equation}
In these equations the Jost matrices are taken at the complex resonant energy.

\begin{table}
\begin{center}
\begin{tabular}{|c|c|c|c|}
\hline
$E_r$\,(keV) & $\Gamma$\,(keV) & $\Gamma_{n\alpha}$\,(keV) &
$\Gamma_{dt}$\,(keV)\\
\hline
9.1 & 9.0 & 7.4 & 1.6\\
\hline
50.2 & 46.3 & 29.1 & 17.2\\
\hline
62.1 & 92.8 & 76.6 & 16.2\\
\hline
\end{tabular}
\end{center}
\caption{\sf
Parameters of the resonances found for the coupled channels 1 and 2 (see
Table~\ref{table.channels}).
}
\label{table.resonances}
\end{table}

For the coupled channels 1 and 2, we found three resonances on the
non-physical Riemann sheet, $(--)_0$, with the parameters given in Table
\ref{table.resonances}.
As was expected from our qualitative reasoning, the partial widths for the
$dt$-channel turned out to be significantly smaller than the corresponding
widths for the $n\alpha$-channel. In the (uncoupled) channels 3 and 4, no
resonances were found.

By scanning all four principal sheets of the Riemann surface within a distance
of $\sim50$\,keV around the central energy $E_0$, we found several $S$-matrix
poles on each of the sheets. These poles are listed in
Table \ref{table.allpoles}.

\begin{table}
\begin{center}
\begin{tabular}{|c|c|c|c|}
\hline
\multicolumn{4}{|c|}{\sffamily $S$-matrix poles (keV)}\\
\hline
$(++)_0$ & $(-+)_0$ & $(+-)_0$ & $(--)_0$ \\
\hline
$9.0-i4.6$   & $9.1-i4.6$   & $9.1-i4.5$   & $9.1-i4.5$ \\
$43.8-i33.8$ & $57.1-i26.5$ & $47.8-i38.3$ & $50.2-i23.2$ \\
$55.5-i23.9$ & $73.6-i26.3$ & $51.2-i22.0$ & $62.1-i46.4$ \\
\hline
$9.0+i4.6$   & $9.1+i4.6$   & $33.3+i22.6$ & $43.0+i57.5$ \\
$43.8+i33.8$ & $57.1+i26.5$ & $42.5+i57.3$ & $48.5+i33.2$ \\
$55.5+i23.9$ & $73.6+i26.3$ & $62.3+i19.4$ & \\
\hline
\end{tabular}
\end{center}
\caption{\sf
Poles of the two-channel $S$-matrix on all the principal sheets of the Riemann
surface (see Fig.~\ref{fig.sheets_2ch_Coulomb}).
The energy is counted from the $dt$-threshold.
}
\label{table.allpoles}
\end{table}

The poles that are adjacent to the physical scattering energies and 
therefore may have influence on physical observables, are those given in the 
left bottom and right upper blocks of this Table. It is seen that the poles on 
the sheets $(++)_0$ and $(-+)_0$ have the ``mirror'' partners while this is not 
the case for the sheets $(--)_0$ and $(+-)_0$. This is a consequence of the 
fact that the Coulomb interaction breaks down the ``mirror'' symmetry and in 
our problem the Coulomb potential is only present in the second channel. A more 
detailed explanation can be found in the Appendix~\ref{Appendix.Coulomb}.

There was a possibility that some of the poles might be spurious, i.e. appearing 
as a result of the truncation of the Taylor series (\ref{A.Taylor}) and 
(\ref{B.Taylor}). The reason for possible appearance of the spurious poles can 
be understood using the following simple example. Consider the exponential 
function $e^z$ that does not have zeros anywhere on the complex $z$-plane. 
However, if it is approximated by a finite number of the Taylor terms, it 
becomes a polynomial that has $N$ zeros, where $N$ is its highest power. 
Apparently, all such zeros are spurious.

In our case, each $S$-matrix pole
corresponds to a zero of the Jost matrix determinant (see Eqs. (\ref{detfzero})
and (\ref{Smatrix})), and the Jost matrix obtained by fitting the experimental
data, involves the polynomials (\ref{A.Taylor}) and (\ref{B.Taylor}). Of course
there are many other (non-polynomial) factors in the Jost matrices, but a
possibility that some spurious zeros may appear cannot be excluded.
In a more simple (non-Coulomb) problem the existence of such spurious zeros was
demonstrated in Ref~\cite{my.effRange}.

As with any spurious solutions, there is a simple recipe to identify them: they
are unstable and drastically change as a result of any small changes in the
parameters of the problem. In our problem, there is a parameter whose choice is
rather arbitrary and the results should not depend on such a choice. This is
the center $E_0$ of the Taylor expansions (\ref{A.Taylor}) and (\ref{B.Taylor}).
These expansions are valid within a circle around $E_0$ and therefore it should
be chosen somewhere not far from the energy where it is expected to find
resonances. Of course for different choices of $E_0$ the fitting parameters
$\bm{a}^{(m)}$ and $\bm{b}^{(m)}$ are different, but the non-spurious zeros of
the Jost matrix determinant must be the same.

In order to check if the poles listed in Table \ref{table.allpoles}
are non-spurious, we repeated the fit with
$E_0=40$\,keV and $E_0=60$\,keV (in additional to our original fit with
$E_0=50$\,keV). The quality of these fits are the same as shown in Figs.
\ref{fig.fit11}, \ref{fig.fit12}, \ref{fig.fit21}, and \ref{fig.fit22}. The
corresponding parameters $\bm{a}^{(m)}$ and $\bm{b}^{(m)}$ are given in Table
\ref{table.parameters12}. As is
seen, the fitting parameters are indeed very different for the three choices of
$E_0$. However, the positions of all the poles listed in Table
\ref{table.allpoles}, turned out to be stable with respect to the
variations of $E_0$ (the changes are in the fourth
digit). On these grounds, we conclude that all these poles are meaningful, i.e.
non-spurious.

\begin{table}
\begin{center}
\begin{tabular}{|c|c|c|c|c|}
\hline
\sffamily resonances (keV) & \sffamily shadow poles (keV) &
\sffamily sheet & \sffamily Ref. & \sffamily year\\
\hline
$48.10-i37.08$ & $78.94-i8.26$ & $(??)$ & \cite{Brown} & 1987\\
\hline
$46.97-i37.1$ & $81.57-i3.64$ & $(-+)$ & \cite{PRL59} & 1987\\
\hline
$47-i36$ & $77+i14$ & $(?-)$ & \cite{Karnakov} & 1990\\
\hline
$47-i37$ & $82-i3.4$ & $(-+)$ & \cite{Bogdanova} & 1991\\
\hline
$48-i41$ & $88-i21$ & $(-+)$ & \cite{betan} & 2018\\
\hline
\end{tabular}
\end{center}
\caption{\sf
The resonant and the shadow $S$-matrix poles in the state $\frac32^+$ near the
$dt$-threshold, reported in several publications. The question marks mean that
the authors did not give the corresponding signatures (identities) of the
Riemann sheets.
}
\label{table.comparison}
\end{table}

In Table~\ref{table.comparison}, for the purpose of comparison, we list the
poles reported in several other publications~\cite{Brown, PRL59, Karnakov,
Bogdanova, betan}. 
As is seen, we found much more
poles. However, it should be noted that we do not contradict the 
previous findings. Indeed, Our resonance at the energy $\sim(50-i23)$\,keV on 
the sheet $(--)_0$ is not far from the pole $\sim(47-i37)$\,keV (on the same 
sheet) reported by the other authors (see Table~\ref{table.comparison}). And one 
of our ``shadow'' poles, namely, at $\sim(74-i26)$\,keV on the sheet $(-+)_0$, 
is not far from the corresponding ``shadow'' poles obtained in the other 
publications that are cited in Table~\ref{table.comparison}.
Of course, there are some differences, but they are not too big. We therefore 
can say that we have located the same poles but at somewhat shifted points. 

As to the other poles we found, there are several possible reasons why they 
were missed previously. It might be that nobody actually searched for poles in 
the locations that were considered as being ``inappropriate''. In fact there is 
no rigorous mathematical theory that would tell us where the poles of a 
multi-channel $S$-matrix can be and where cannot. Yet another possible reason 
for the differences  between the Tables \ref{table.allpoles} and 
\ref{table.comparison} may be rooted in the different analytic structure of the 
functions used for the analytic continuation. As an example of such a 
difference it can be mentioned the following. In the $R$-matrix theory the 
$S$-matrix is a ratio of two matrices, both of which have poles on the real 
axis. Within our approach, the $S$-matrix is also a ratio of two (Jost) 
matrices, but they do not have poles at real energies.

In order to pinpoint the actual cause of the difference, it is needed to 
perform a mathematically rigorous analysis of the analytic structure of the 
$R$-matrix, of the $S$-matrix derived from it, and of the topology of the 
Riemann surface on which they are defined. To the best of our knowledge, nobody 
ever did such an analysis and many of the $R$-matrix properties on the Riemann 
surface are tacitly assumed.

Some erroneous assumptions might lead to wrong conclusions.
For example, in Fig.~5 of Ref.~\cite{Bogdanova} it is assumed that one can
continuously move from the lower half of the sheet $(-+)_0$ (denoted in
\cite{Bogdanova} as $U_{(2)}$) across the connected rims of the cut to the upper
half of the sheet $(+-)_0$ (denoted as $U_{(1)}$). According to our analysis
(see Fig.~\ref{fig.sheets_2ch_Coulomb}), this is wrong. It would be correct 
without the Coulomb interaction. But in the presence of the Coulomb forces the 
number of the Riemann sheets becomes infinite and some of their 
interconnections become different.

Moreover, as we show in the Appendix~\ref{Appendix_symmetry}, the Coulomb
forces break down most of the symmetry relations among the elements of the
$S$-matrix on different sheets of the Riemann surface. As a result, the
reasoning  about the role of
the shadow poles, given by Eden and Taylor in Ref.~\cite{EdenTaylor}, is not
applicable for the charged particles. However, most of the authors base their
conclusions on the so-called ``generalized unitarity'' of
Ref.~\cite{EdenTaylor}.

The elastic and inelastic scattering can only be affected by the poles that are
close to the real axis where the physical and non-physical
sheets $(++)_0$ and $(--)_0$ are connected. All the other poles are too far.
Indeed, it is necessary to go around one or more branch points in order
to reach the scattering energies from them.

\begin{figure}
\centerline{\epsfig{file=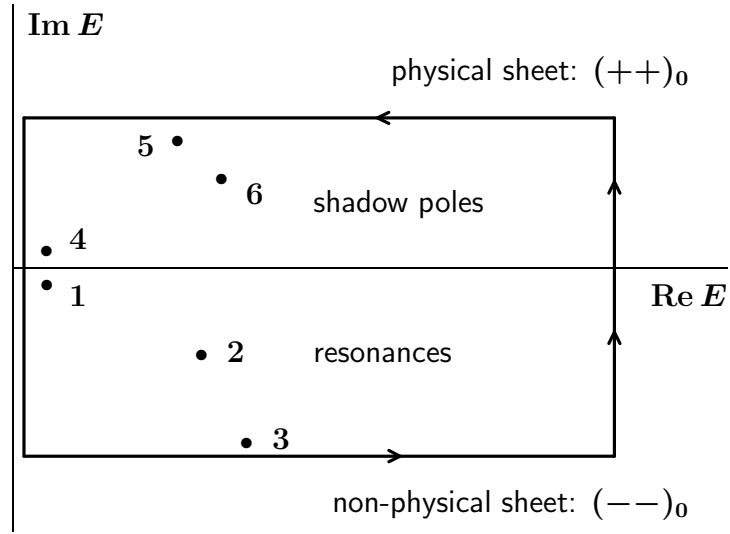}}
\caption{\sf
The $S$-matrix poles located around the real axis of the scattering energies
that connects the physical and non-physical sheets of the Riemann surface.
The complex values of the corresponding energies can be found in
Tables \ref{table.allpoles} and \ref{table.Residues}. The numerical labels of
the poles are used as the corresponding references in Fig.~\ref{fig.MLexcl}.
The rectangular contour is used for the integration in the Mittag-Leffler sum
(\ref{MittagLeffler}).
}
\label{fig.6poles}
\end{figure}

Therefore among the 23 poles given in Table \ref{table.allpoles}, there are
six poles which may influence the collision process. As is
shown in Fig.~\ref{fig.6poles}, they are on both sides of the real axis
which is between $(++)_0$ and $(--)_0$.
The poles with negative imaginary parts are the usual resonances, and those
above the real axis are the shadow poles.

In order to learn about the importance of each of these six poles for the
scattering process, we apply the method used in Ref.~\cite{our_J.Q.Chem.}. This
method is based on the Mittag-Leffler theorem (see, for example,
Ref.~\cite{MLbook}), which allows one to split a meromorphic function in the
pole and the non-singular (background) terms. This is what is described next.

\subsection{Contributions from individual poles}
\label{sec.Mittag-Leffler}
Let us consider the closed rectangular contour shown in Fig.~\ref{fig.6poles},
which goes around the poles nearest to the real scattering energies, i.e. the 
poles on the interconnected sheets $(++)_0$ and $(--)_0$ of the Riemann 
surface.
If $E$ is a point inside this contour (we choose it on the real axis), then
according to the residue theorem, we have
\begin{equation}
\label{Res_theorem}
   \oint\frac{\bm{S}(\zeta)}{\zeta-E}d\zeta=
   2\pi i\bm{S}(E)+2\pi i\sum_{j=1}^L
   \frac{\mathrm{Res}[\bm{S},E_j]}{E_j-E}\ ,
\end{equation}
where $E_j$ are the poles enclosed by the contour ($L=6$). Therefore
\begin{equation}
\label{MittagLeffler}
   \bm{S}(E)=\sum_{j=1}^L
   \frac{\mathrm{Res}[\bm{S},E_j]}{E-E_j}+
   \frac{1}{2\pi i}\oint\frac{\bm{S}(\zeta)}{\zeta-E}d\zeta\ ,
\end{equation}
which is a particular formulation of the more general Mittag-Leffler theorem.
In Eq.~(\ref{MittagLeffler}) the $S$-matrix $\bm{S}(E)$ on the real axis is
written as a sum of the contributions from each pole and a background integral.

\begin{table}
\begin{center}
\begin{tabular}{|c|c|c|c|}
\hline
\sffamily sheet & \sffamily pole: $E$\,(keV) &
\sffamily $\mathrm{Res}[S_{nn'},E]$\,(keV) & $n,n'$ \\
\hline
\hline
\multirow{12}{*}{$(--)_0$} &
\multirow{4}{*}{$9.1-i4.5$} & $0.023681-i0.036357$   & 1,1 \\
&&           $0.0035344-i0.0036706$ & 1,2 \\
&&           $0.47319+i0.087305$    & 2,1 \\
&&           $0.052651+i0.020520$   & 2,2 \\
\cline{2-4}		
&
\multirow{4}{*}{$50.2-i23.2$} & $-0.51181+i0.65675$ &  1,1 \\
&           & $-0.28605+i 0.28668$ &  1,2 \\
&           & $-9.2422+i 5.9593$ &  2,1 \\
&           & $-4.8154+i2.3285$   &  2,2 \\
\cline{2-4}	
&
\multirow{4}{*}{$62.1-i46.4$} & $1.7887-i14.855$    &  1,1 \\
&           &  $2.2774+i 1.5480$  &  1,2 \\
&           & $36.478+i 49.327$   &  2,1 \\
&           & $-11.147+i1.7942$   &  2,2 \\
\hline
\hline
\multirow{12}{*}{$(++)_0$} &
\multirow{4}{*}{$9.0+i4.6$} & $-0.024456+i 0.039395$ & 1,1 \\
&          & $-0.0073593+i 0.0027477$ & 1,2 \\
&          & $-0.092407+i 0.73479$ & 2,1 \\
&          & $-0.088505+i 0.088928$ & 2,2 \\
\cline{2-4}		
&
\multirow{4}{*}{$43.8+i33.8$} & $-12.589+i 14.992$ &  1,1 \\
&           & $1.7237-i 0.63052$ &  1,2 \\
&           & $39.860-i 30.546$ &  2,1 \\
&           & $-4.6670+i 0.62084$ &  2,2 \\
\cline{2-4}	
&
\multirow{4}{*}{$55.5+i23.9$} & $1.1959+i 2.0737$ &  1,1 \\
&           & $0.56551+i 0.17848$ &  1,2 \\
&           & $14.951+i 1.5362$ &  2,1 \\
&           & $2.9874-i 2.2224$ &  2,2 \\
\hline
\end{tabular}
\end{center}
\caption{\sf
Poles of the two-channel $S$-matrix and the corresponding residues of its
elements in the domains of the Riemann sheets $(--)_0$ and $(++)_0$
adjacent to the axis of the real scattering energies (see
Fig.~\ref{fig.sheets_2ch_Coulomb}).
The energy is counted from the $dt$-threshold.
}
\label{table.Residues}
\end{table}

After fitting the experimental data, we obtain the analytic fomulae for the
Jost matrices and hence for the $S$-matrix. This allows us to (numerically)
calculate the background integral for any given scattering energy $E$. The
poles $E_j$ are known. We assume that all the poles are simple. Therefore the
residues at them can be found by numerical differentiation of the determinant of
the Jost matrix. Indeed,
\begin{equation}
\label{SmatrixExplicit}
    \bm{S}=\bm{f}^{(\rm out)}\left(
    \begin{array}{cc}
    f^{(\rm in)}_{22} & -f^{(\rm in)}_{12} \\[3mm]
    -f^{(\rm in)}_{21} & f^{(\rm in)}_{11}
    \end{array}\right)
    \frac{1}{\det \bm{f}^{(\rm in)}}
\end{equation}
and thus
\begin{equation}
\label{ResidueExplicit}
    {\rm Res}\,[\bm{S},E]=\bm{f}^{(\rm out)}(E)\left(
    \begin{array}{cc}
    f^{(\rm in)}_{22}(E) & -f^{(\rm in)}_{12}(E) \\[3mm]
    -f^{(\rm in)}_{21}(E) & f^{(\rm in)}_{11}(E)
    \end{array}\right)
    \left[\frac{d}{dE}\det \bm{f}^{(\rm in)}(E)\right]^{-1}\ ,
\end{equation}
where
\begin{equation}
\label{DerivativeExplicit}
   \frac{d}{dE}\det \bm{f}^{(\rm in)}(E)\approx
   \frac{\det \bm{f}^{(\rm in)}(E+\epsilon)-\det \bm{f}^{(\rm in)}(E-\epsilon)}
   {2\epsilon}\ .
\end{equation}
In our calculations, we used $\epsilon=1$\,eV which gave the
accuracy of at least 5 digits. Thus calculated residues for the six poles are
given in Table \ref{table.Residues}. With these residues and with the
numerically calculated background integral (the contour was extended up to
200\,keV and on each line-segment of the contour it were used 400
Gauss-Legendre integration points) we obtained (as it should be) exactly the
same cross sections that are shown in Figs.
\ref{fig.fit11}, \ref{fig.fit12}, \ref{fig.fit21}, and \ref{fig.fit22}.
This is a kind of cross-check of our calculations. Moreover this fact is an
additional proof that there is no other poles (apart from the six poles shown
in Fig.~\ref{fig.6poles}) inside the contour.

\begin{figure}
\centerline{\epsfig{file=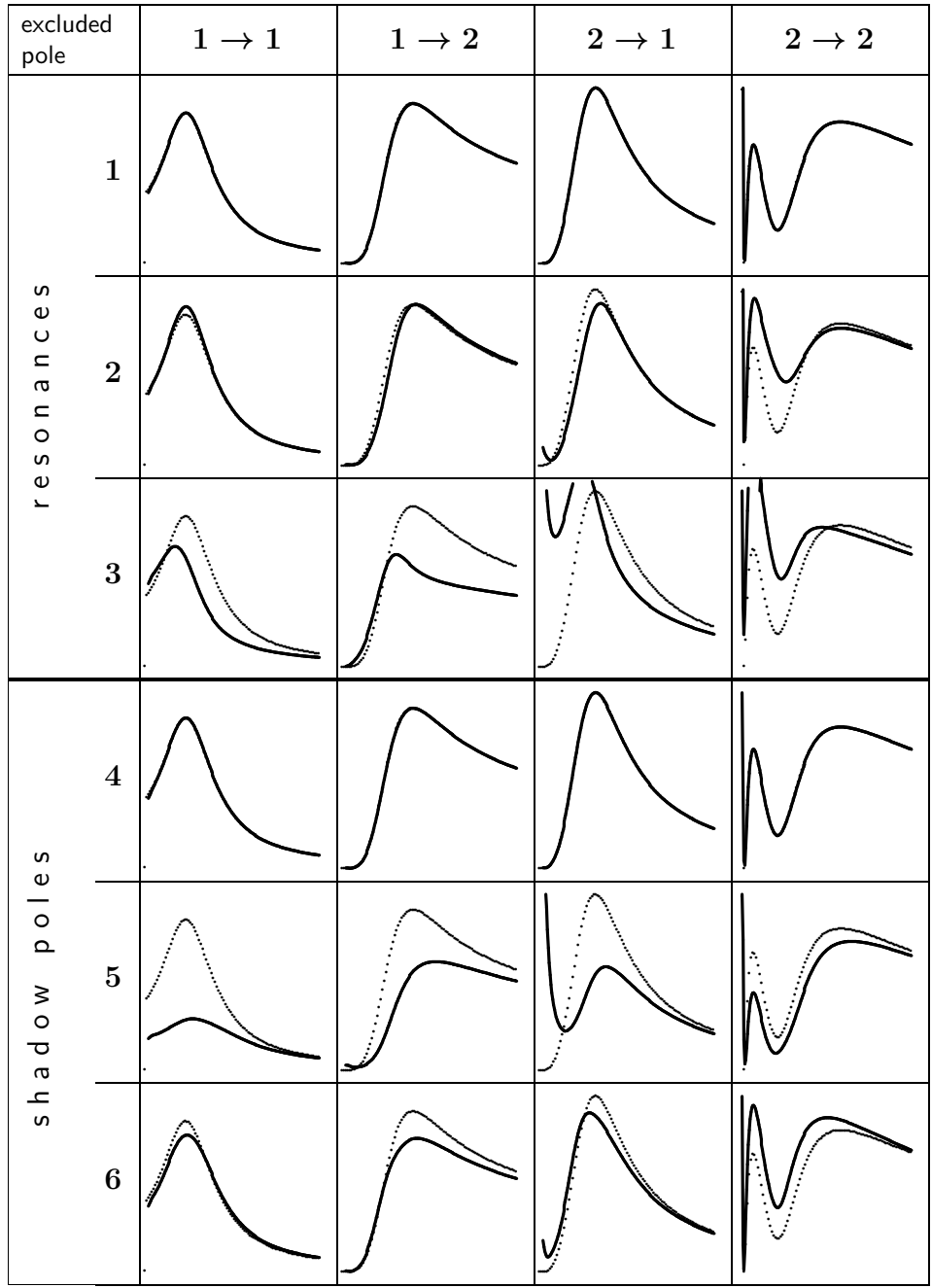}}
\caption{\sf
Partial cross sections (the curves) for the inter-channel transitions $n\to
n'$, when a
single pole is excluded from the Mittag-Leffler sum (\ref{MittagLeffler}). The
channels are labeled as in Table~\ref{table.channels}. The poles are numbered
as is shown in Fig.~\ref{fig.6poles}. The dots are the corresponding
experimental (i.e. the $R$-matrix) cross sections.
}
\label{fig.MLexcl}
\end{figure}

Now we can omit some of the poles from the sum (\ref{MittagLeffler})
and see how this affects the partial cross sections. The results of such an
analysis (where we excluded only one pole at a time) are shown in
Fig.~\ref{fig.MLexcl}. The curves show the cross sections when one pole is
excluded. The dots are the experimental data (i.e. the $R$-matrix cross
sections).

It is seen that the first resonance as well as the first
shadow pole (poles 1 and 4, respectively) practically do not affect the cross
sections at all (their effect is in the fourth digit). This means that these two
poles can be safely ignored. The exclusion of the second resonance (pole number
2) makes noticeable changes only in the elastic $dt\to dt$ scattering.
Meanwhile, the third resonance (pole number 3) is important for all four
(elastic and inelastic) processes. Among the three shadow poles, the second one
(pole number 5) is the most important. Its contribution to all four processes
is comparable with the effect of the second resonance. The third shadow pole
(pole number 6) makes just ``cosmetic'' changes to all the partial cross
sections.

\begin{figure}
\centerline{\epsfig{file=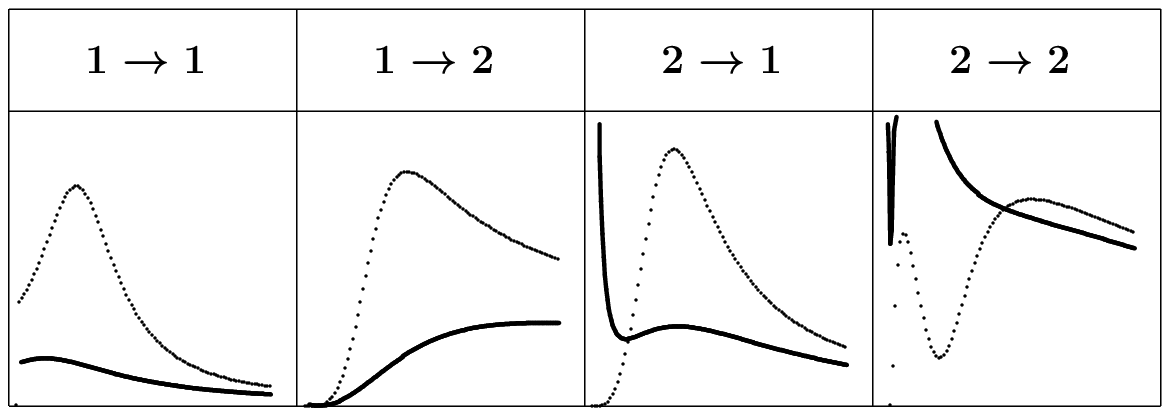}}
\caption{\sf
Partial cross sections for the inter-channel transitions $n\to n'$, when all
six poles shown in Fig.~\protect{\ref{fig.6poles}} are excluded from the
Mittag-Leffler sum (\ref{MittagLeffler}). The dots are the corresponding
experimental (i.e. the $R$-matrix) cross sections.
}
\label{fig.Nopoles}
\end{figure}

It is also interesting to know what happens if we exlude all the pole terms
from the Mittag-Leffler expansion (\ref{MittagLeffler}) and only leave the
background integral. Fig.~\ref{fig.Nopoles} shows the result of such an
exclusion. As is seen, without the resonant and the shadow poles all the cross
sections are far from the experimental points. In the channels $2\to1$ and
$2\to2$ the background integral is responsible for the sharp peaks near the
threshold energy.

\section{Summary and conclusion}
\label{sec.conclusion}
As was stated in the Introduction, the main question that we wanted to answer
in the present work was about the nature of
${}^5\mathrm{He}^*(\frac32^+)$-resonance. Using an available $R$-matrix fit as
our input (experimental) data, we constructed the corresponding Jost-matrices
that have proper analytic structure 
and are defined on the Riemann surface with the proper topology (where both the 
square-root and logarithmic branching are present).
Exploring their behaviour on various sheets
of the Riemann surface, we located 23 poles of the $S$-matrix. Only 6 of these
poles turned out to be close enough to the axis of the real scattering energies
and therefore only these 6 poles could influence the observable quantities. The
set of these 6 poles consists of 3 resonances and 3 shadow poles.

Using the Mittag-Leffler representation, we isolated the
individual contributions to the $S$-matrix from all the resonances and the 
shadow poles. In this way it was established that the energy dependencies of the
partial cross sections near the $dt$-threshold are mainly determined by the two
resonant poles,
\begin{eqnarray*}
    &&\left(50.2-\frac{i}{2}46.3\right)\,\mathrm{keV}\ ,
    \quad \Gamma_{n\alpha}=29.1\,\mathrm{keV}\ ,
    \quad \Gamma_{dt}=17.2\,\mathrm{keV}\ ,\\
    &&\left(62.1-\frac{i}{2}92.8\right)\,\mathrm{keV}\ ,
    \quad \Gamma_{n\alpha}=76.6\,\mathrm{keV}\ ,
    \quad \Gamma_{dt}=16.2\,\mathrm{keV}\ ,
\end{eqnarray*}
and two shadow poles at
$\left(43.8+i33.8\right)\,\mathrm{keV}$ and
$\left(55.5+i23.9\right)\,\mathrm{keV}$ on the principal physical sheet
$(++)_0$.
The contribution of the other pair of the resonance and its shadow (at around
$\sim9$\,keV) is negligible and can be ignored. The non-resonant background
scattering is responsible for the threshold cusps in the channels
$dt\to n\alpha$ and $dt\to dt$.

\appendixpage
\appendix
\section{Symmetry properties of the Jost matrices}
\label{Appendix_symmetry}
The matrices $\bm{f}^{\mathrm{(in)}}$ and $\bm{f}^{\mathrm{(out)}}$ are related
to each other at different points of the Riemann surface, i.e. they obey certain
symmetry rules. Since the unknown matrices $\bm{A}$ and $\bm{B}$ in the
semi-analytic representations
(\ref{multi.matrixelements}) and (\ref{multichannel.JostMatr.Jostfact}) are the
same on all the sheets, the symmetry relations among the Jost matrices are
determined by the explicitly given factors that undergo certain changes when we
replace $k_n$ with $-k_n$ or with $k_n^*$. For the sake of clarity, we firstly
consider the simplified representation (\ref{multichannel.JostMatr.Jostfact})
for neutral particles. Then, using the more general
Eq.~(\ref{multi.matrixelements}), we show that the Coulomb forces break some of
the symmetries.

\subsection{Neutral particles}
\label{Appendix.neutral}
The energy variable $E$ determines all the channel momenta, but not completely.
There is the freedom to choose their signs. Different signs of them place the
corresponding point on this or that sheet of the Riemann surface. In order to
specify which sheet of the surface the point belongs to, it is convenient to
replace the notation $\bm{f}^{\rm(in/out)}(E)$ with
$\bm{f}^{\rm(in/out)}(k_1,k_2,\dots,k_N)$. To simplify the formulae, we will
write them for $N=2$.

\begin{figure}[ht!]
\centerline{\epsfig{file=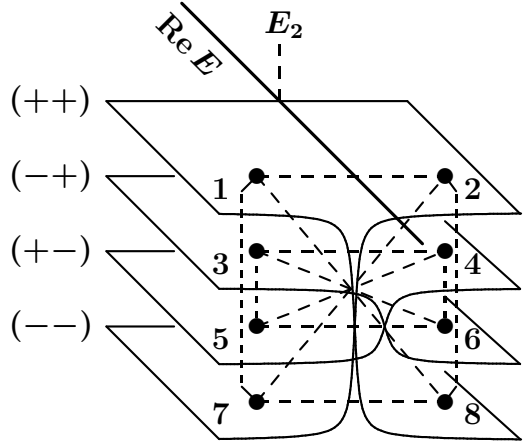}}
\caption{\sf
Fragment of the Riemann surface for a two-channel problem of neutral particles. 
The interconnections of the sheets are shown for the energies above the second 
threshold, $E_2$. The points with the 
odd numbers 
correspond to the same energy $E$, while the points with the even numbers 
correspond to the complex conjugate energy $E^*$.
The dashed lines connect the points at which the values of the Jost 
matrices are related according to 
Eqs.~(\protect{\ref{sign_symm}}), 
(\protect{\ref{multichannel_short_finSwartz}}), and 
(\protect{\ref{multichannel_short_FinCombSym}}).
}
\label{fig.multi_Riemann_symmetries}
\end{figure}

Replacing $(k_1,k_2)$ with $(-k_1,-k_2)$ in
Eq.~(\ref{multichannel.JostMatr.Jostfact}), we see that the factorized momenta
generate a common factor $(-1)^{\ell_m+\ell_n}=(-1)^{\ell_m-\ell_n}$ and the
sign between the two terms, i.e the $(\mp)$, is changed to the opposite,
$(\pm)$. This means that
\begin{equation}
\label{multichannel_short_FinFoutSym}
      f^{\rm (in/out)}_{mn}(-k_1,-k_2)=
      (-1)^{\ell_m+\ell_n}f^{\rm (out/in)}_{mn}(k_1,k_2)\ .
\end{equation}
The change $(k_1,k_2)\to(-k_1,-k_2)$ is equivalent to moving to a different
sheet of the Riemann surface, namely, $(++)\leftrightarrow(--)$ or
$(-+)\leftrightarrow(+-)$. Such a transition is done to a point that is above
or below the initial location on a vertical line that corresponds to the same
energy. If the parity is conserving, then $(-1)^{\ell_m+\ell_n}=1$ and
Eq.~(\ref{multichannel_short_FinFoutSym}) relates the whole matrices,
\begin{equation}
\label{sign_symm}
      \bm{f}^{\rm (in/out)}(-k_1,-k_2)=
      \bm{f}^{\rm (out/in)}(k_1,k_2)\ .
\end{equation}
This can be called the ``vertical'' symmetry. It is schematically shown in 
Fig.~\ref{fig.multi_Riemann_symmetries}, where
$$
      \bm{f}^{\rm (in/out)}(1)=
      \bm{f}^{\rm (out/in)}(7)\ ,
      \qquad
      \bm{f}^{\rm (in/out)}(3)=
      \bm{f}^{\rm (out/in)}(5)\ ,
$$
$$
      \bm{f}^{\rm (in/out)}(2)=
      \bm{f}^{\rm (out/in)}(8)\ ,
      \qquad
      \bm{f}^{\rm (in/out)}(4)=
      \bm{f}^{\rm (out/in)}(6)\ .
$$

\begin{figure}
\centerline{\epsfig{file=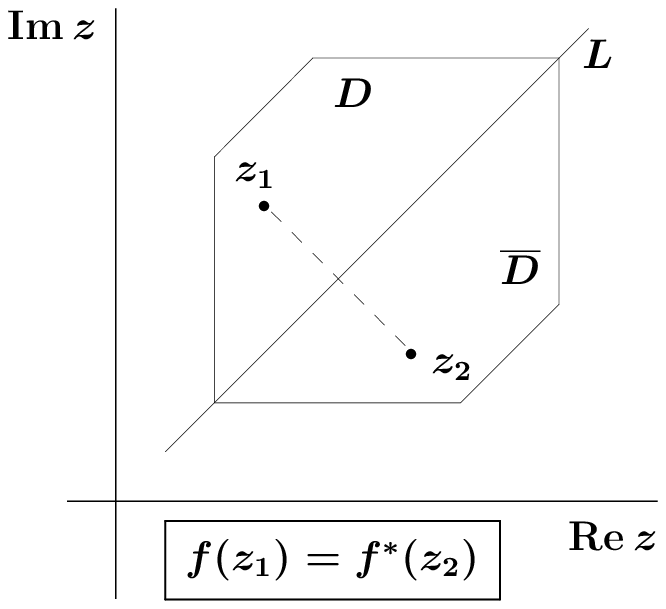}}
\caption{\sf
Illustration of the Schwartz reflection principle.
}
\label{II.RS.fig.Schwartz}
\end{figure}

One more symmetry property can be derived using the Schwartz reflection
principle, which reads (see, for example Ref.~\cite{boas.cmplx}): If a function
$f(z)$ is analytic in a domain $D$ that includes a line segment $L$, and $f(z)$
is real on $L$, then this function can be analytically continued across $L$ into
the domain $\overline{D}$, which is a mirror reflection of $D$ relative to the
line $L$, and the values of $f(z)$ at any two points symmetric with respect to
$L$ are complex conjugates of each other. It is graphically illustrated in Fig.
\ref{II.RS.fig.Schwartz}.

The functions (matrices) $\bm{A}(E)$ and $\bm{B}(E)$ are real on the real
axis\cite{our.MultiCh, our_Coulomb}.  Therefore
\begin{equation}
\label{multichannel_short_Schwartz}
    \bm{A}(E^*)=\bm{A}^*(E)\ ,
    \qquad
    \bm{B}(E^*)=\bm{B}^*(E)\ .
\end{equation}
The change $E\to E^*$ is equivalent to $(k_1,k_2)\to(k_1^*,k_2^*)$,
which implies that
\begin{eqnarray*}
   f_{mn}^{\rm(in/out)}(E^*)
   &=&
   \left(\frac{k_n^{\ell_n+1}}{k_m^{\ell_m+1}}\right)^*
   A_{mn}(E^*)-
   i\left(k_m^{\ell_m}k_n^{\ell_n+1}\right)^*
   B_{mn}(E^*)\,=\,\left[f_{mn}^{\rm(out/in)}(E)\right]^*\ ,
\end{eqnarray*}
and thus
\begin{equation}
\label{multichannel_short_finSwartz}
   \bm{f}^{\rm(in/out)}(E^*)=\left[\bm{f}^{\rm(out/in)}(E)\right]^*\ .
\end{equation}
Apparently, when replacing the energy variable $E$ with its complex conjugate 
partner, $E^*$, the imaginary parts of 
all the channel momenta change their signs. As a result, the point $E^*$ is on 
the other side of the real axis and on a different sheet of the Riemann 
surface, which can symbolically be written as 
follows:
\begin{equation}
\label{milti.symm.EEstar}
   E\to E^*\quad\Rightarrow\quad
   (\mathrm{sign}_1,\mathrm{sign}_2)\to
   (-\mathrm{sign}_1,-\mathrm{sign}_2)\ .
\end{equation}
The symmetry  (\ref{multichannel_short_finSwartz}) is valid for each pair of the 
sheets having the opposite signatures as in Eq.~(\ref{milti.symm.EEstar}),
and can be called the ``diagonal'' symmetry. It is schematically shown by the 
diagonal dashed lines in Fig.~\ref{fig.multi_Riemann_symmetries}, where
\begin{equation}
\label{diag_symm_example}
\begin{array}{rcllrcl}
      \bm{f}^{\rm (in/out)}(1) &=&
      \bm{f}^{\rm (out/in)*}(8)&,&
      \qquad
      \bm{f}^{\rm (in/out)}(3) &=&
      \bm{f}^{\rm (out/in)*}(6)\ ,\\[3mm]
      \bm{f}^{\rm (in/out)}(5) &=&
      \bm{f}^{\rm (out/in)*}(4)&,&
      \qquad
      \bm{f}^{\rm (in/out)}(7) &=&
      \bm{f}^{\rm (out/in)*}(2)\ .
\end{array}
\end{equation}
We can combine the ``vertical'' and ``diagonal'' symmetries stated by Eqs.
(\ref{sign_symm}) and (\ref{multichannel_short_finSwartz}), which gives
\begin{equation}
\label{multichannel_short_FinCombSym}
      \bm{f}^{\rm (in/out)}(k_1,k_2)=
      \left[\bm{f}^{\rm (in/out)}(-k_1^*,-k_2^*)\right]^*\ .
\end{equation}
When changing the sign, $k_n\to-k_n$, we change the sign of $\mathrm{Im}\,k_n$.
And the complex conjugation brings the same sign back. Therefore the two points
$(k_1,k_2)$ and $(-k_1^*,-k_2^*)$ are on the same
Riemann sheet, but on the opposite sides of the cut. As a result, we obtain the
symmetry relation along the horizontal lines shown in 
Fig.~\ref{fig.multi_Riemann_symmetries}, where 
$$
      \bm{f}^{\rm (in/out)}(1)=
      \bm{f}^{\rm ((in/out)*}(2)\ ,
      \qquad
      \bm{f}^{\rm (in/out)}(3)=
      \bm{f}^{\rm (in/out)*}(4)\ ,
$$
$$
      \bm{f}^{\rm (in/out)}(5)=
      \bm{f}^{\rm (in/out)*}(6)\ ,
      \qquad
      \bm{f}^{\rm (in/out)}(7)=
      \bm{f}^{\rm (in/out)*}(8)\ .
$$
This is known as the ``mirror symmetry''. Apparently, if
there is a spectral points at the energy $E$, for example, if
$\det \bm{f}^{\rm(in)}(7)=0$,
then there is a ``mirror'' resonance at the point $8$ (see 
Fig.~\ref{fig.multi_Riemann_symmetries}).
In the two-channel case, the ``mirror'' resonances can be
between the thresholds $E_1$ and $E_2$ on the sheet $(-+)$ on the other side of
the cut, and for possible resonances above the threshold $E_2$ they are on
the sheet $(--)$ also on the opposite side of the cut. In order to reach
the real axis of the physical sheet $(++)$ from such ``mirror'' resonances,
we have to go around one and two branch points, respectively. This means that
they are far away from the scattering energies and thus their influence on the
observable quntities is very weak, if any.

Using the definition (\ref{Smatrix}) and the symmetries
of the Jost matrices shown in Fig.~\ref{fig.multi_Riemann_symmetries}, it is 
easy to derive the corresponding symmetry-relations for the $S$-matrix. For 
example,
$$
      \bm{S}(1)=
      \left[\bm{S}^{-1}(8)\right]^*\ ,
      \qquad
      \bm{S}(3)=
      \left[\bm{S}^{-1}(6)\right]^*\ ,
$$
$$
      \bm{S}(5)=
      \left[\bm{S}^{-1}(4)\right]^*\ ,
      \qquad
      \bm{S}(7)=
      \left[\bm{S}^{-1}(2)\right]^*\ .
$$
It should not be forgotten that this is only valid when
$(-1)^{\ell_m+\ell_n}=1$, i.e. for the potentials that conserve the parity.

\subsection{Interaction involving a Coulomb tail}
\label{Appendix.Coulomb}
If the multi-channel potential has a Coulomb tail at least in one of the
channels, some of the symmetries  symbolically depicted in 
Fig.~\ref{fig.multi_Riemann_symmetries}, are broken. 
This can be shown by analysing the analytic structure given by 
Eq.~(\ref{multi.matrixelements}), which we can re-write in the matrix form as 
follows,
\begin{equation}
\label{multi.Finout_structure}
   \bm{f}^{(\mathrm{in/out})}=
   \bm{Q}^{(\pm)}\left[
   \bm{D}^{-1}\bm{A}\bm{D}-(\bm{M}\pm i)
   \bm{K}^{-1}\bm{D}\bm{B}\bm{D}\right]\ ,   
\end{equation}
where the (depending on the energy) diagonal matrices $\bm{Q}^{(\pm)}(E)$,
$\bm{D}(E)$, $\bm{M}(E)$, and $\bm{K}(E)$ are defined as (see 
Ref.\cite{our_Coulomb}):
\begin{equation}
\label{multi.defQ}
   \bm{Q}^{(\pm)} =
   \operatorname{diag}\left\{
   \frac{e^{\pi\eta_1/2}\ell_1!}{\Gamma(\ell_1+1\pm i\eta_1)},
   \frac{e^{\pi\eta_2/2}\ell_2!}{\Gamma(\ell_2+1\pm i\eta_2)},\dots,
   \frac{e^{\pi\eta_N/2}\ell_N!}{\Gamma(\ell_N+1\pm i\eta_N)}\right\}\ ,
\end{equation}

\begin{equation}
\label{multi.defD}
   \bm{D} =
   \operatorname{diag}\left\{
   C_{\ell_1}(\eta_1)k_1^{\ell_1+1},
   C_{\ell_2}(\eta_2)k_2^{\ell_2+1},\dots,
   C_{\ell_N}(\eta_N)k_N^{\ell_N+1}\right\}\ ,
\end{equation}

\begin{equation}
   \bm{M} =
   \operatorname{diag}\left\{
   \frac{2\eta_1h(\eta_1)}{C_0^2(\eta_1)},
   \frac{2\eta_2h(\eta_2)}{C_0^2(\eta_2)},
   \dots,
   \frac{2\eta_Nh(\eta_N)}{C_0^2(\eta_N)}\right\}\ ,
\end{equation}

\begin{equation}
\label{multi.chmom}
   \bm{K}=\operatorname{diag}\left\{k_1,k_2,\dots,k_N\right\}\ .
\end{equation}
First of all, let us consider what happens if we change the sign of a channel 
momentum $k_n$. 
The sign can be changed by the exponential 
factor
\begin{equation}
\label{multi.C.signchange}
   k_n\ \to\ -k_n\ :
   \qquad
   e^{i\pi(2j+1)}k_n\ ,\qquad j=\pm1,\pm2,\dots
\end{equation}
For the channel $n$ without a Coulomb potential, all values of $j$ in such a
factor are equivalent. However, if there are Coulomb forces in the 
channel $n$, then the functions involved in the matrices of
Eq.~(\ref{multi.Finout_structure}), undergo the following changes: 
\begin{equation}
\label{multi.Coulomb.sign.change.matr}
   k_n\ \to\ e^{i\pi(2j+1)}k_n\ \Rightarrow\ \left\{
   \begin{array}{lcl}
   \eta_n\to -\eta_n\ ,\\[3mm]
   C_{\ell_n}(\eta_n) &\to& e^{\pi\eta_n}C_{\ell_n}(\eta_n)\ ,\\[3mm]
   h(\eta_n) &\to& h(\eta_n)+ i\pi(2j+1)\ .
   \end{array}\right.
\end{equation}
Apparently, after the change of the sign of a channel momentum, these functions 
have different values and thus the relation
(\ref{multichannel_short_FinFoutSym}), generally speaking, cannot be valid 
anymore. In other words, the ``vertical'' symmetry is broken by the Coulomb 
forces. 

It is easy to see that the ``diagonal'' symmetry that follows from the Schwartz 
reflection principle, remains valid in the presence of the Coulomb potential. 
The ``diagonal'' transition $E\to E^*$ moves the point to the other side of the 
real axis.  Both the original point $E$ and the new point $E^*$ can be on any 
of the Riemann sheets. It is clear that the complex conjugation of the energy 
results in the complex conjugation of all the channel momenta and thus of all 
the Sommerfeld parameters,
$$
   E\to E^*\ \Rightarrow\quad k_n\to k_n^*\ ,\quad \eta_n\to\eta_n^*\ ,
   \qquad \forall\ n\ .
$$
The transition $k_n\to k_n^*$ changes the sign of $\mathrm{Im}\,k_n$. This 
means that the point $E^*$ is on a different sheet of the Riemann surface. 
Actually, all the signs in the sheet-label $(\pm,\pm,\dots)_{m_1m_2,\dots}$ are 
changing to the opposite, as is given by Eq.~(\ref{milti.symm.EEstar}). 
Apparently, the logarithmic indices, $m_1m_2,\dots$, 
change to the opposite as well. This can be seen from 
Eq.~(\ref{Logarithm_ch_momenta}) that defines them.

It is clear that
   $\bm{Q}^{(\pm)}(E^*)=\left[\bm{Q}^{(\mp)}(E)\right]^*$,
   $\bm{D}(E^*)=\left[\bm{D}(E)\right]^*$,
   $\bm{M}(E^*)=\left[\bm{M}(E)\right]^*$,   
   and
   $\bm{K}(E^*)=\left[\bm{K}(E)\right]^*$. 
Taking into account the Schwartz relations (\ref{multichannel_short_Schwartz}), 
we therefore arrive at the same symmetry 
property (\ref{multichannel_short_finSwartz}),
\begin{equation}
\label{multichannel_Coul_finSwartz}
   \bm{f}^{\rm(in/out)}(E^*_{m_1m_2,\dots})=
   \left[\bm{f}^{\rm(out/in)}(E_{-m_1,-m_2,\dots})\right]^*\ ,
\end{equation}
where the subscripts (logarithmic indices $m_1,m_2,\dots$) of the energy 
variable specify the sheet of the Riemann surface. 
The corresponding relations 
for the $S$-matrix  can be obtained as follows:
\begin{eqnarray}
\nonumber
   \bm{S}(E^*_{m_1m_2,\dots})
   &=&
   \bm{f}^{\rm(out)}(E^*_{m_1m_2,\dots})
   \left[\bm{f}^{\rm(in)}(E^*_{m_1m_2,\dots})\right]^{-1}\\[3mm]
\nonumber
   &=&
   \bm{f}^{\rm(in)*}(E_{-m_1,-m_2,\dots})
   \left[\bm{f}^{\rm(out)*}(E_{-m_1,-m_2,\dots})\right]^{-1}\\[3mm]
\nonumber
   &=&
   \left\{
   \bm{f}^{\rm(out)*}(E_{-m_1,-m_2,\dots})
   \left[\bm{f}^{\rm(in)*}(E_{-m_1,-m_2,\dots})\right]^{-1}
   \right\}^{-1}\\[3mm]
\label{multi.C.S.diagonalsymm}   
   &=&
   \left[\bm{S}^*(E_{-m_1,-m_2,\dots})\right]^{-1}\ .
\end{eqnarray}
It should not be forgotten that the transition 
$E^*_{m_1m_2,\dots}\to E_{-m_1,-m_2,\dots}$ moves the point to a different 
sheet of the Riemann surface, where all the labels are changing to the 
opposite, for example,
$$
   (++-)_{1,0,-1}\ \to\ (--+)_{-1,0,1}\ .
$$
In particular, the relation (\ref{multi.C.S.diagonalsymm}) means that a 
resonance pole of the $S$ matrix corresponds to a zero of its determinant at 
the symmetric point on the physical sheet of the Riemann surface.

\begin{figure}
\centerline{\epsfig{file=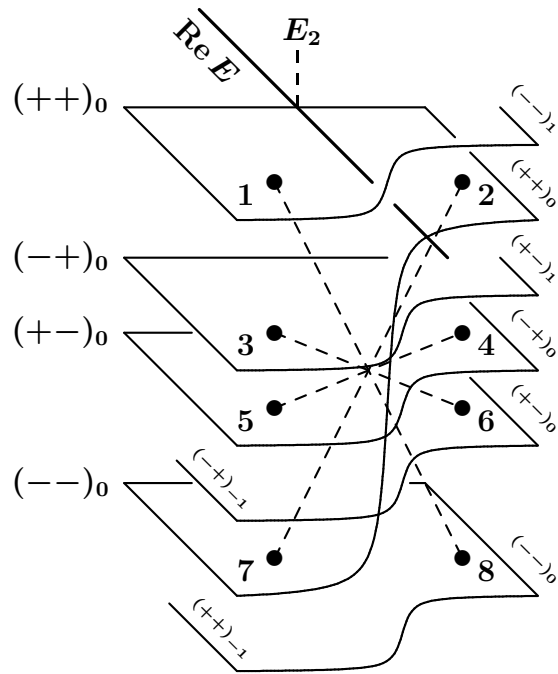}}
\caption{\sf
Fragment of the Riemann surface shown in 
Fig.~\protect{\ref{fig.sheets_2ch_Coulomb}}(b). The points with the odd numbers 
correspond to the same energy $E$, while the points with the even numbers 
correspond to the complex conjugate energy $E^*$.
The four dashed lines connect the points at which the values of the Jost 
matrices are related according to 
Eq.~(\protect{\ref{multichannel_Coul_finSwartz}}).
}
\label{fig.sheets_2ch_Coulomb_fragment_1m}
\end{figure}

For a two-channel problem, the diagonal relations 
(\ref{multichannel_Coul_finSwartz}) are illustrated in 
Fig.~\ref{fig.sheets_2ch_Coulomb_fragment_1m}. Of course there are much more 
(actually infinite number of) ``diagonal'' relations of the type 
(\ref{multichannel_Coul_finSwartz}). 
Fig.~\ref{fig.sheets_2ch_Coulomb_fragment_1m} 
only shows the symmetric points on the principal branch of the Riemann surface. 
The other sheets (with non-zero logarithmic indices) have no practical 
influence 
on the observables and thus can be safely ignored. In this figure, the points 
labeled by the odd numbers, are on the same vertical line and thus correspond 
to the same energy $E$. The even numbers label the symmetric complex conjugate 
energies, $E^*$. The diagonal dashed lines correspond to the four 
relations given by Eq.~(\ref{diag_symm_example}).
It is instructive to compare the 
relations depicted in Fig.~\ref{fig.sheets_2ch_Coulomb_fragment_1m} with the 
corresponding relations for the neutral particles given in 
Fig.~\ref{fig.multi_Riemann_symmetries}.

Since there are no ``vertical'' symmetries for the charged particles, the 
``horizontal'' symmetries (that result from the combination of the 
``diagonal'' and ``vertical'' ones) are also broken. In particular, this means 
that there are no ``mirror'' poles of the $S$-matrix, if the charges are 
non-zero. For example, if the point $7$ in 
Fig.~\ref{fig.sheets_2ch_Coulomb_fragment_1m} is a resonance, then the point 
$8$ could be its ``mirror'' partner, but in contrast to the neutral-particle 
case (see Fig.~\ref{fig.multi_Riemann_symmetries}), the points $7$ and $8$ are 
not related by a ``horizontal'' symmetry.

When doing the numerical calculations, we compared the Jost matrices at 
all the symmetric points shown in 
Fig.~\ref{fig.sheets_2ch_Coulomb_fragment_1m}. In such a 
comparison, we observed the ``diagonal'' symmetry and did not see any other 
relations among the values of the Jost matrices. This is in accordance with the 
above analysis. However, we also (numerically) found an inexplicable relation, 
namely, the zeros of $\det\bm{f}^{\rm(in)}$ at the energies $E$ on the sheets 
$(++)_0$ and $(-+)_0$ have the ``mirror'' zeros at the corresponding points 
$E^*$ on the same sheets, as if it were the ``horizontal'' symmetry for these 
sheets. The calculations at the nearby points show that there is no such a 
``horizontal'' symmetry, but the zeros are still symmetric. It looks like the 
determinant is factorized in a symmetric and non-symmetric functions, but we 
could not proof this analytically.

\end{document}